\documentclass[graybox, x11names, table]{svmult}

\usepackage{mdframed}
\usepackage{graphicx}
\renewcommand\thechapter{\arabic{chapter}}
\renewcommand\thesection{\arabic{section}}

\usepackage[utf8]{inputenc}

\usepackage{type1cm} 

\usepackage{array}

\usepackage{makeidx}  
\usepackage{graphicx} 
\usepackage{multicol} 
\usepackage[bottom]{footmisc}

\usepackage{newtxtext} %
\usepackage[varvw]{newtxmath} 

\usepackage{lipsum}

\usepackage{titlesec}

\usepackage{enumitem}

\usepackage{comment}

\usepackage{todonotes}

\usepackage{wrapfig}

\usepackage[center,font=small,skip=2pt]{caption}

\usepackage{graphicx}
\usepackage{subcaption}

\usepackage{xcolor}

\usepackage[bookmarks=false]{hyperref}
\usepackage{soul}
\usepackage{tablefootnote}
\hypersetup{
  colorlinks = {true},
  linkcolor = {blue}, 
  linkbordercolor = {white},
  citecolor = {blue},
  urlcolor = {blue}
}

\usepackage{tabularray}
\usepackage{color}

\usepackage{multirow} 

\usepackage{doi}
\usepackage{orcidlink}

\makeindex  

\titlespacing\section{2pt}{12pt plus 4pt minus 2pt}{3pt plus 2pt minus 2pt}
\titlespacing\subsection{2pt}{12pt plus 4pt minus 2pt}{3pt plus 2pt minus 2pt}
\titlespacing\subsubsection{2pt}{12pt plus 4pt minus 2pt}{3pt plus 2pt minus 2pt}

\setlist[itemize]{noitemsep}

\begin{document}

\title{Towards a Consensual Definition for\\
Smart Tourism and Smart Tourism Tools}

\author{
António Galvão\inst{{1,2} \orcidlink{0000-0002-6566-9114}}
\and
Fernando Brito e Abreu \inst{{1} \orcidlink{0000-0002-9086-4122}}
\and
João Joanaz de Melo \inst{{2} \orcidlink{0000-0003-4138-0014}}
}

\institute{
1) Instituto Universitário de Lisboa (ISCTE-IUL), ISTAR, Lisboa, Portugal,\\ \email{(antonio.galvao, fba)@iscte-iul.pt}\\
2) NOVA University Lisbon - School of Science and Technology, CENSE, Caparica, Portugal,\\ \email{amg13172@campus.fct.unl.pt, jjm@fct.unl.pt}
}

\maketitle

\abstract{\\
Smart tourism (ST) stems from the concepts of e-tourism - focused on the digitalization of processes within the tourism industry, and digital tourism – also considering the digitalization within the tourist experience. The earlier ST references found regard ST Destinations and emerge from the development of Smart Cities.
\\
Our initial literature review on the ST concept and Smart Tourism Tools (STT) revealed significant research uncertainties: ST is poorly defined and frequently linked to the concept of Smart Cities; different authors have different, sometimes contradictory, views on the goals of ST; STT claims are often only based on technological aspects, and their “smartness” is difficult to evaluate; often the term “Smart” describes developments fueled by cutting-edge technologies, which lose that status after a few years.
\\
This chapter is part of the ongoing initiative to build an online observatory that provides a comprehensive view of STTs' offerings in Europe, known as the \textbf{European STT Observatory}. To achieve this, the observatory requires methodologies and tools to evaluate 'smartness' based on a sound definition of ST and STT, while also being able to adapt to technological advancements. In this chapter, we present the results of a participatory approach where we invited ST experts from around the world to help us achieve this level of soundness. Our goal is to make a valuable contribution to the discussion on the definition of ST and STT.
}

\section{Introduction}
\label{sec:introduction}

The \href{https://www.resetting.eu/}{RESETTING project}\footnote{Relaunching European smart and SustainablE Tourism models Through digitalization and INnovative technoloGies} aims at facilitating the shift towards more robust, circular, and eco-friendly business practices for tourism companies across Europe. Within its scope we are building an online observatory to provide a broad view of STTs' offer in Europe, the \textbf{European STT Observatory}. This observatory requires methods and tools to assess "smartness", based on a sound definition of ST and STT and able to cope with technological evolution. This chapter is scoped in this effort.

To correctly understand ST (offer), a preliminary literature review on the ST concept and on STT, and a broad online search on STT, were performed. These early efforts revealed two significant research gaps or rather research uncertainties.

Although there is a common agreement in the academia that Smart Tourism (ST) refers to the use of technologies to improve visitors' and local's experiences, while enabling sustainability goals, different authors consider different key aspects, from techno-, through tourist- or business-, to destination- centered definitions. Additionally, the hype around "Smart" has led to the misuse of the concept, known as "SmartWashing" \cite{Desdemoustier2019}. STT are frequently just claims based on technological aspects provided by developers; the “smartness” of such technologies is often difficult to evaluate.

Additionally, the term "Smart" is used to describe developments fueled by cutting-edge technologies, and since technology development is in constant evolution, STTs are a moving target where a “Smart” offer quickly becomes a “Dumb” offer.

To populate the European STT Observatory we need (a) strong definition(s) for ST(T). Additionally, this endeavor will enable us to establish an STT smartness index to gauge the intelligence of the European STT market. In addition, this effort will allow us to create an STT smartness index to measure the intelligence of the European STT supply. This chapter outlines our ongoing efforts to achieve these goals, focusing on our scientific and collaborative approach to reach a unanimous definition of STT.

Nowadays travel has became more accessible, convenient, and enjoyable for millions of people around the world. However, there is a digital divide between large enterprises and small and medium-sized enterprises (SMEs) in terms of their stake in the tourism industry \cite{MinghettiVandBuhalisD2010, Reverte2020}. The tourism industry's adoption of digitalization, innovation, and new technologies necessitates the use of STT. However, tourism SMEs lag behind in their ability to capitalize on the opportunity of a digital transformation of their core business due to financial and technical limitations, as well as a lack of awareness of existing STT. This is the driving force behind the creation of the European STT Observatory.

Thus our research is framed in the following research questions:
\begin{itemize}
    \item \textbf{RQ1:} What are the main criteria and methods for evaluating the smartness level of STT and comparing them across different domains and regions?
    \item \textbf{RQ2:} How can the participatory approach involving ST experts help to reach a consensual definition of ST and STT and to create a robust taxonomy for ST?
    \item \textbf{RQ3:} How can the European STT Observatory provide a comprehensive and up-to-date overview of the STT offer in Europe?
    \item \textbf{RQ4:} What are the benefits and challenges of using STT for tourism SMEs and how can the European STT Observatory help to reduce the digital gap between SMEs and large enterprises?
\end{itemize}

\section{Literature Review}
\label{LiteratureReview}
\subsection{Smart Tourism: Evolution}
\label{subsec:(Smart)TourismFoundations}

The word ``tourist" is believed to have originated in the 18th century, derived from the French word ``tour," meaning ``turn" or ``trip" and referred to individuals who embarked on a journey or circuit, often for leisure or educational purposes\footnote{Still nowadays the Merriam-Webster Dictionary defines tourism as \textit{``the practice of traveling for recreation"} \cite{merriam-websterTourism}}. Others claim that the etymology of the term tourism is much older, tracing back to the ancient Greek word for a tool used in describing a circle. In a sense, tourism is a journey that starts and ends at the same place, home \cite{Leiper1983}. Nomadic lifestyles have accompanied humanity since its earliest development and are responsible for the widespread distribution of human beings. Although early human migrations were probably motivated by a combination of factors, including climate change \cite{liu2016}, population growth, and competition for resources, the advent of religions and culture, embodied in prehistoric monuments, suggests that early humans visited these sites as a form of primitive tourism.

However, the modern tourism industry began in the 19th century with the development of transportation infrastructure such as railways and steamships, and in the 1950s, with the widespread ownership of the automobile and the arrival of charter flights, national and international tourism took off  \cite{Christou2022, mowforth2015, SharpleyR2022}.

In the pre-internet era, tourism suppliers had to rely on intermediaries for their distribution functions. The emergence of Computer Reservation Systems (CRSs) in the 1970s and Global Distribution Systems (GDSs) in the late 1990s facilitated the intermediation process, but it was the development of the Internet in the 1990s, leveraged by Information and Communication Technologies (ICTs) by providing tools for marketing, customer relationship management, yield management, quality control and innovation, that changed both business practices and strategies and ultimately reshaped the tourism industry \cite{Buhalis2008}. As ICT evolved, new forms of tourism were shaped. The impact of new technologies is so great that it calls into question the very definition of tourism. We live in an era in which tourism has gone beyond the limits of a circular route between home and destination, and today we can travel without leaving home (virtual tourism \cite{Verma2022}), visit different times or experience parallel realities in the areas where we travel (metaverse tourism and extended reality \cite{YangAndWang2023}), or even aspire to visit places outside our own atmosphere (space tourism \cite{Reddy2012}).With the unprecedented exponential growth we are currently witnessing thanks to recent developments in AI, who knows what the future holds. It is in this context that ST is developing.

The evolution of smart tourism has been a transformative process, marked by the integration of advanced technologies into the tourism industry. The next paragraphs trace the journey from the early stages of e-tourism to the current era of smart tourism.

Tourism was one of the first sectors to digitalize business processes on a global scale, bringing flight and hotel booking online to become a digital pioneer. As ICT became a global phenomenon, tourism was a consistent early adopter of new technologies and platforms \cite{Gössling2021}.

The early digitization of tourism was coined as ``e-tourism" and was focused on digitizing processes within the tourism industry, changing the way organizations distributed their tourism products in the market \cite{Buhalis2002}. This led to the development of online booking systems, e-ticketing, and other digital tools that made it easier for travelers to plan and book their trips online. Soon after, the new concept of ``Digital Tourism" was proposed \cite{Zambonelli_etal_2001}. It is a broader term that encompasses e-tourism but also includes the use of digital tools by tourists to prepare, organize, control, and enjoy their travel experience.

In the early 2000s, the rise of mobile technology and smartphones paved the way for the development of ST. According to \cite{Li2017} the term ``Smart Tourism" can be traced back to 2000 when Gordon Phillips defined it as a sustainable approach to planning, developing, operating and marketing tourism products and businesses \cite{Phillips2000}\footnote{This presentation in \href{http://slideshare.net}{SlideShare} could not be found, so this claim could not be verified}. In his opinion, ST is shaped by two types of techniques: 1) smart demand and the use of management techniques that are capable of managing demand and access; 2) smart marketing techniques that can be used to target the proper customer segments to deliver appropriate messages.

Some preliminary research on STT can be found in \cite{YiLong_etal_2008}. According to \cite{Wang2013}, the earliest reference to Smart Tourism Destinations (STD) was coined in the ``STD initiative'', promoted by China's State Council of Chinese Central Government in 2009, in relation to a platform on which information relating to tourists activities, the consumption of tourism products, and the status of tourism resources could be instantly integrated and then provided to tourists, enterprises, and organizations through a variety of end-user devices.

The concept of ST became widespread in the turn of the 2010s, inspired by the development of Smart Cities, and supported by new technologies such as mobile apps, augmented reality (AR), virtual reality (VR), and the Internet of Things (IoT) to provide tourists with personalized and immersive experiences \cite{BuhalisAndAditya2014, Wang2013}.

E-tourism, Digital Tourism, and ST came in succession chronologically, portraying the revolutionary changes brought to the tourism industry over the past few decades, leveraged by the power of technology. However, these concepts are often used interchangeably, as they all involve the use of technology to enhance and facilitate tourism activities \cite{Kononova2020}. Yet other technology-inspired designations for tourism have been used in scientific literature, such as ``Mobile tourism (M-tourism)", ``Intelligent Tourism", ``Tourism 4.0" and ``Virtual Tourism" \cite{Kononova2020}.

\subsection{Smart Tourism: a fuzzy concept}
\label{subsec:FuzzySmartTourism}
Despite all the hype around ST there is a lack of consensus on the definition of ST \cite{Celdrán-Bernabéu2018, Gretzel_etal_2015, Dejan_etal_2021, Li2017, Chen-Kuo_etal_2020, Rodrigues2022, Randhir2020, Jahanyan_etal_2022, Shafiee2021, Zhang2016}. Several authors argue that the development of ST is hindered by the lack of such a definition \cite{Celdrán-Bernabéu2018, Gretzel_etal_2015, Randhir2020, Jahanyan_etal_2022, Shafiee2021}. However, producing that definition is an endeavor jeopardized by the inherent complexity of the ST ecosystem, excellently discussed in \cite{Gretzel2015}.

Celdrán-Bernabéu et al. \cite{Celdrán-Bernabéu2018} reason that future research aimed at the theoretical-conceptual development and critical analysis of ST should fill the existing gap between knowledge and operational development of ST. Additionally, they find some regionality associated to the terminology. In Western countries, ST is not a core strategy of tourism development but is based on its contribution to sustainable development and the relationship between tourists and destinations. In East Asia, ST actions focus on policies that promote the development of technological infrastructure, while in Europe most initiatives are identified with Smart City projects that have favored the emergence of the smart destination approach.

Other authors point out the weaknesses resulting from this lack of clarity. For instance, Rodrigues et al. \cite{Rodrigues2022} point out that \textit{“some existing smart tourism strategies seem to use both smart and sustainability concepts in a marginal, propagandist way, more similar to a marketing point of view. This might be due to theoretical inconsistencies, which need to be addressed in future studies, defining a reliable basis for fully understanding the smart approach in tourism.”} Li et al. \cite{Li2017} argue that because of such practices there is the risk that the concept of ST could be abandoned.

To highlight the different views on ST, we extracted the key concepts underlying the definitions of ST compiled by Kononova et al. \cite{Kononova2020}. The results were grouped in classes representing the broader considerations in those definitions (Table \ref{table:keySTdefConcepts}).

\begin{table}[htb]
\centering
\caption{Key concepts found in Smart Tourism definition}
\begin{tabular}{|p{0.48\linewidth} | p{0.48\linewidth}|}
\hline
\rowcolor[HTML]{ececec} 
\textbf{Technology and Innovation}        & \textbf{Sustainability}   \\ \hline
• Maximizing environmental, cultural, social and economic values through IT        & • Intelligent, meaningful and sustainable connections   \\
• Mobile digital connectivity     & • Deep civic engagement    \\
• Combined model of tourism industry and innovative technology      & • Clean, eco-friendly, ethical and high-quality services   \\ \cline{2-2} 
• Data accumulation with technological means       & \cellcolor[HTML]{ececec}\textbf{Tourist Experience}   \\ \cline{2-2} 
• Technology-driven innovation    & • Individual tourist support system   \\
• Constant and systematic use of smart elements        & • Ubiquitous tour information service   \\
• Device generated big data for monitoring behavior, tourism management and marketing      & • Automatic provision of suitable and precise services  \\
• Potential replacement of human labor through digital technologies      & • Integrated efforts to find innovative ways for data accumulation and aggregation or use \\
• Evolutionary development of traditional tourism and e-tourism      & • Interaction with a more comfortable environment for both locals and tourists \\
• Sensors, data mining, positioning technology, SNS and social network technology        & • Creation of additional travel value for the tourist   \\ \cline{2-2} 
• Products using technological components       & \cellcolor[HTML]{ececec}\textbf{Smart Cities}   \\ \cline{2-2} 
• Privacy preserving, context awareness, recommender systems, social media, IoT, user experience, real-time, user modeling, augmented reality and big data & • Inspired by the idea of smart cities    \\ \hline
\end{tabular}
\label{table:keySTdefConcepts}
\end{table}

\subsection{Related works}

\label{subsubsec:STTapplications_relatedWork}
One interesting approach in refining ST definition is brought by Chen et al. \cite{Chen_etal_2021}. The authors argue that existing research on ST focuses on technological application and the opinions of users and suppliers, lacking discussion from the perspective of academic experts. Furthermore, they claim that most studies focus only on a single aspect of people, the planning process, or the technological components of ST. In this context, they interviewed 11 ST scholars to examine their views on the role of and interactions among three key components (people, process, and technology) in current ST development. The main findings (per aspect) are summarized below.

\underline{\textbf{People – key stakeholders}}

Informants generally agreed that several stakeholders are involved in the development of ST, including tourists, suppliers, governments, destination marketing organizations, service providers, system providers, the technology itself, and residents. However, informants revealed different levels of importance among all these stakeholders.
The development of ST is closely linked to that of the smart city and affects the quality of life of residents who should have a say in the planning process.

\underline{\textbf{Process – ST planning process}}

The ST planning process requires the collaboration and effort of various stakeholders, including the tourism board, government, and sectors involved. Cooperation between academia and industry is essential to enable a strategic relationship. Fostering innovation is critical to the success of ST planning. The implementation of a ST related project is an essential component for a destination, as the innovative environment ensures a smooth connection with practices.

\underline{\textbf{Technology – ST technologies}}

ST applications are being implemented in various areas such as entertainment, hotel operation, aviation, payments, experience training, ICT-related devices or applications, medical services, small business operation and transportation. Industries with high interaction between tourists and suppliers are the most relevant for the application of ST. The adaptation of ST technologies occurs depending on changes in tourists’ behaviors and service patterns.

ICT is a prerequisite component for smart ecosystem management, but ST is not just about applying ICT. Community interest and participation are critical to the smooth running of ST destination development. The sharing economy with traditional businesses also raises concerns. The availability of information from end users is an essential mediator at the operational level. ST development can evolve into a major regional or national development strategy, and operational guidance is needed to serve as a reference for the development of ST destinations.

The current study follows a similar approach in reaching out to the scientific community to try to create a consensual definition of ST.

ST operationalization is supported by the development and implementation of STT, supported by different technologies, and operating in diverse domains.
 
Dejan et al. \cite{Dejan_etal_2021} set out to ascertain what is being implemented in self-labeled 'smart' destinations and whether most 'smart' projects actually qualify as such, using a structured operational and innovation technolgy  adoption-based aproach, Smart Actionable Classification Model (SACM). Built on top of Perboli et al. \cite{PERBOLI_etal_2014} taxonomy of Smart City projects, that lacks the evaluation of recent advanced technologies, it is adapted to the context of Smart Tourism, using Buhalis' definition of smart systems in tourism \cite{SmartTourismBuhalis2022} - \textit{“Smart systems use a wide range of networks, connected devices, sensors and algorithms for big data delivery across the IoT.”} – to define the \textbf{Smart Actionable} attributes used to categorize ST projects: (i) \textbf{networked/connected devices} and applications, (ii) coordinated by \textbf{intelligent algorithms}, (iii) based on collected and analyzed information at a \textbf{Big Data} level. They classified 35 ST projects publicly available in Europe, into the following groups, based on their tech type and Smart Actionable:

\bgroup
\def\arraystretch{1.5}
\begin{table}[]
\centering
\caption{Dejan et al.  ST groups}
\begin{tabular}{|p{0.031\linewidth} | p{0.787\linewidth}| p{0.15\linewidth}|}
\hline
\rowcolor[HTML]{ececec} 
\textbf{id} & \textbf{Description} & \textbf{Techs} \\ \hline
S1 & Projects dealing with large amounts of networked and connected data, with the attributes of \textbf{networked/connected} and \textbf{Big Data}. The \ul{use of intelligent algorithms is negligible}. & networks IoT sensors \\ \hline
S2 & Projects connected only by the \textbf{networked/connected} attribute and using to a \ul{lesser extent} \textbf{networked/connected} data. The use of \textbf{intelligent algorithms} is \ul{less frequent} and the \ul{Big Data attribute is absent} (despite most projects using sensors, the collected data is not sufficient to classify them as Big Data)  & sensors mobile apps  \\ \hline
S3 & Projects associated only with the \textbf{Big Data} attribute. They \ul{rarely apply} \textbf{intelligent algorithms} in data processing, and \ul{do not address networked/connected goals}. & IoT AI \\\hline
S4 & Projects that have not been assigned to any Smart Actionable. 
They use significantly fewer individual tech types \tablefootnote{Such projects accounted for as much as 31\% of all projects in the analysis} & mobile apps sensors  \\ \hline
\end{tabular}
\end{table}
\egroup

 The authors found that \textit{``the vast majority of projects branded as ‘smart’ predominantly pursue environmental sustainability goals, but do not feature advanced technology that meets the Smart Actionable attribute criteria, and do not address social sustainability issues to the same extent as the environmental ones''}. This may suggest that there is a gap between the buzz generated by Smartness in tourism and its actual application at the destination level. These findings may also indicate that technocentric metrics alone may not be suitable for evaluating Smartness in the tourism sector. Notably, their study underscores the emphasis on sustainability objectives, with a primary focus on environmental considerations and, to a lesser degree, social aspects. Economic sustainability appears to be of less significance. The authors acknowledge that the projects under analysis, as described publicly, often employ vague technological terminology, particularly, but not exclusively, the term ICT, which impedes the classification of these projects with regards to technological intelligence. They also view their model as a preliminary milestone towards developing a stronger SACM approach, which accounts for additional terminology clarifications, as well as updated technological classification frameworks. Our work achieves a similar outcome, but our projects cannot be classified using the Perboli and colleagues taxonomy. This is because we do not solely focus on smart destination projects, but rather mainly on independent technological solutions that can function without the structured context of a smart city or destination.

Buhalis and Amaranggana \cite{BuhalisandAmaranggana2015} discovered 3 distinct moments in the personalized services expected by tourists in STD:

\begin{enumerate}[leftmargin=1cm,align=left,label={(\arabic*)}]
  \item \emph{Before Trip}: To support the planning phase by giving all the related real-time information based on user profiling, in order to make a more informed decision;
  \item \emph{During Trip}: Enhanced access to real-time information to assist tourists in exploring the destination, direct personalized services, as well as a real-time feedback loop;
  \item \emph{After Trip}: Prolonged engagement to relive the experience, as well as a decent feedback system, allowing tourist to review their holistic tourism experience.
\end{enumerate}

\section{Methodology}
\label{sec:Methodology}

\begin{figure}[htb]
  \centering
  \includegraphics[width=\textwidth]{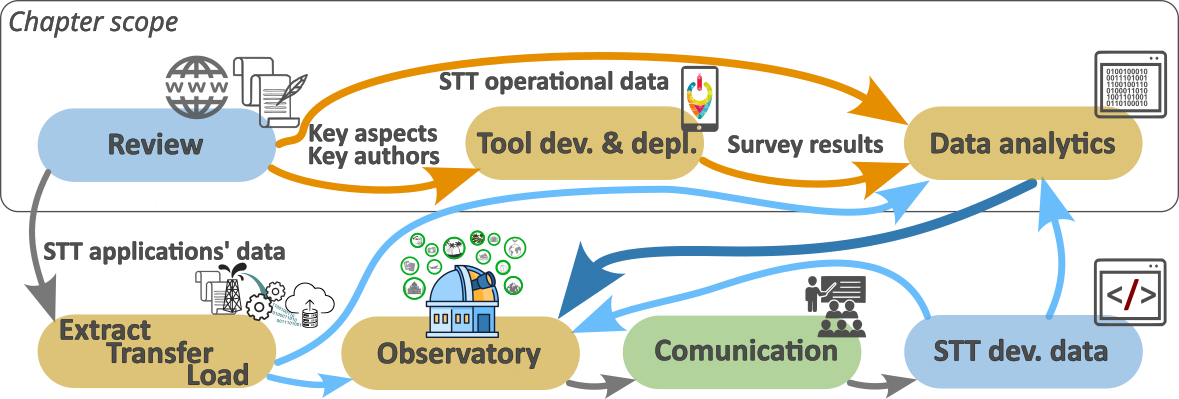}
  \caption{Observatory methodology}
  \label{fig:ObservatoryMethodology}
\end{figure}

In this section, we first present the overall methodology used to build and operate the observatory (Figure \ref{fig:ObservatoryMethodology}, where the \includegraphics[height=1em]{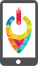} icon represents the RESETTING webapp form). This chapter focuses on the cornerstones for building the observatory:

\begin{itemize}
    \item Our proposed STT taxonomy, based on the operationalization of the studied STT applications and the identified technology domains
    \item The definition of ST and STT, from which we derive the STT Smartness Index, based on our participatory approach
\end{itemize}

The Observatory will be operated and maintained by semi-automated, AI-based Extraction, Transformation and Loading (ETL) processes and expert-in-the-loop, and is available online at \url{https://www.resetting.eu/STTobservatory/}.

\subsection{STT operational taxonomy}
\label{subsubsec:STTapplications_sttOperationTaxonomy}

Several initiatives were developed to collect information on STTs. In Europe, we highlight the Scottish Tourism Toolkit \cite{gidlund2020}, the European Commission report on ST Practices \cite{scholz2022}, Spain's Secretary of State for Tourism Catalogue of Technological Solutions for STD \cite{segittur2022}, and the Spanish Cluster of Innovative Companies for Tourism in the Valencia Community Technological Solutions for Tourism catalog \cite{adestic2023}. We analysed the STTs offered in these catalogs and the descriptions of STTs offered on the Internet, trying to map the areas of operation of STTs, the types of STTs available and the types of technology most commonly used. This allowed us to identify three domains of application for STTs. In this sense, STTs can be:

\begin{itemize}
    \item tools as part of touristic offers;
    \item tools used for the marketing of touristic offers;
    \item tools for managing and operating touristic offers.
\end{itemize}

These domains are not exclusive, i.e. some STTs operate in several domains.

\begin{figure}[htb]
 \centering
 \includegraphics[width=\textwidth]{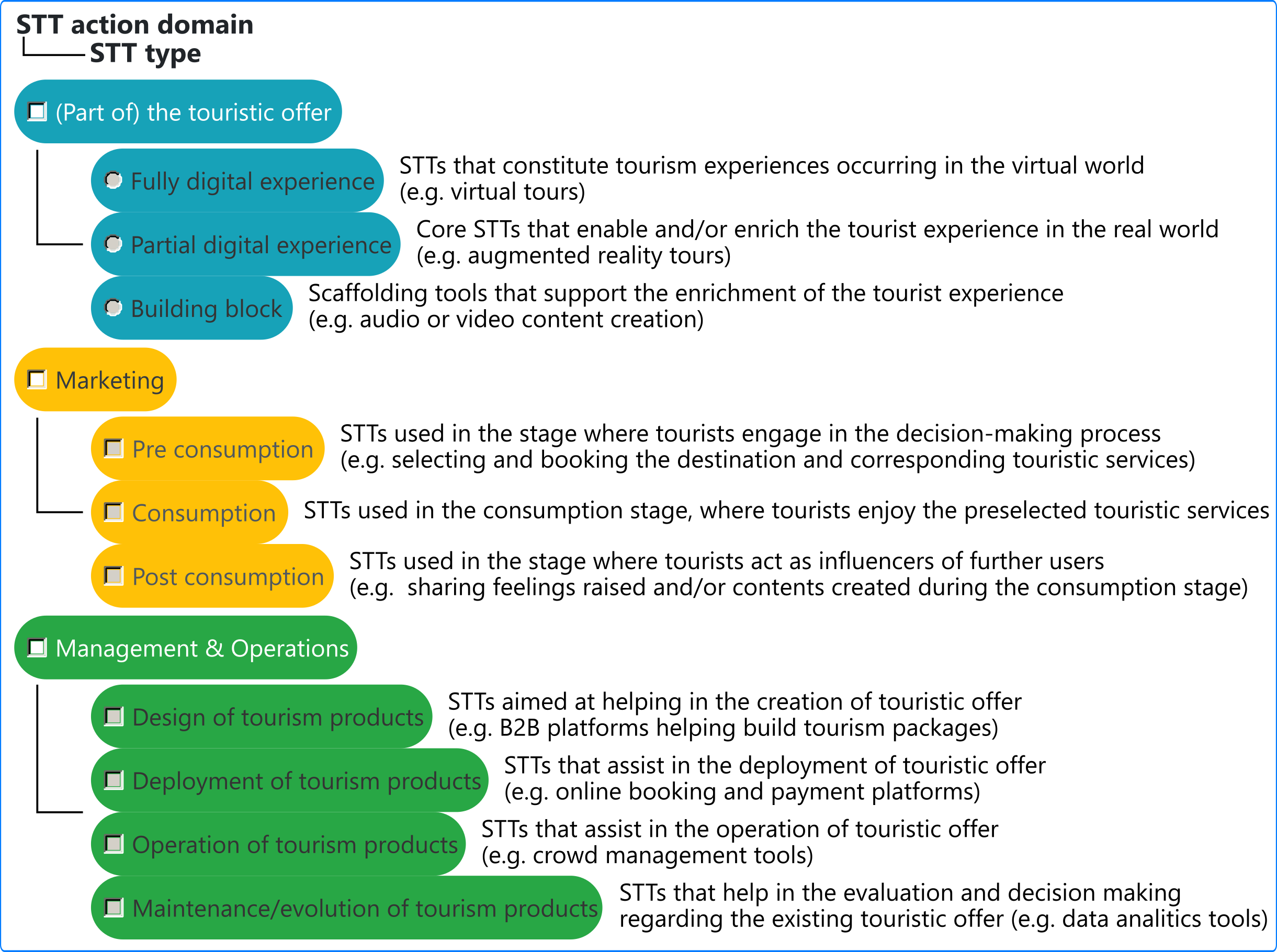}
 \caption{STT action domains and tool types}
 \label{fig:STTactionDomains}
\end{figure}

Regarding the existing types of STT, we found 10 different (non-exclusive) classes (Figure \ref{fig:STTactionDomains}). The main difference between them is not a technological consideration, but rather the differences in the final objectives proposed by the developers and the domains in which they operate.

\begin{figure}[htb]
 \centering
 \includegraphics[width=0.5\textwidth]{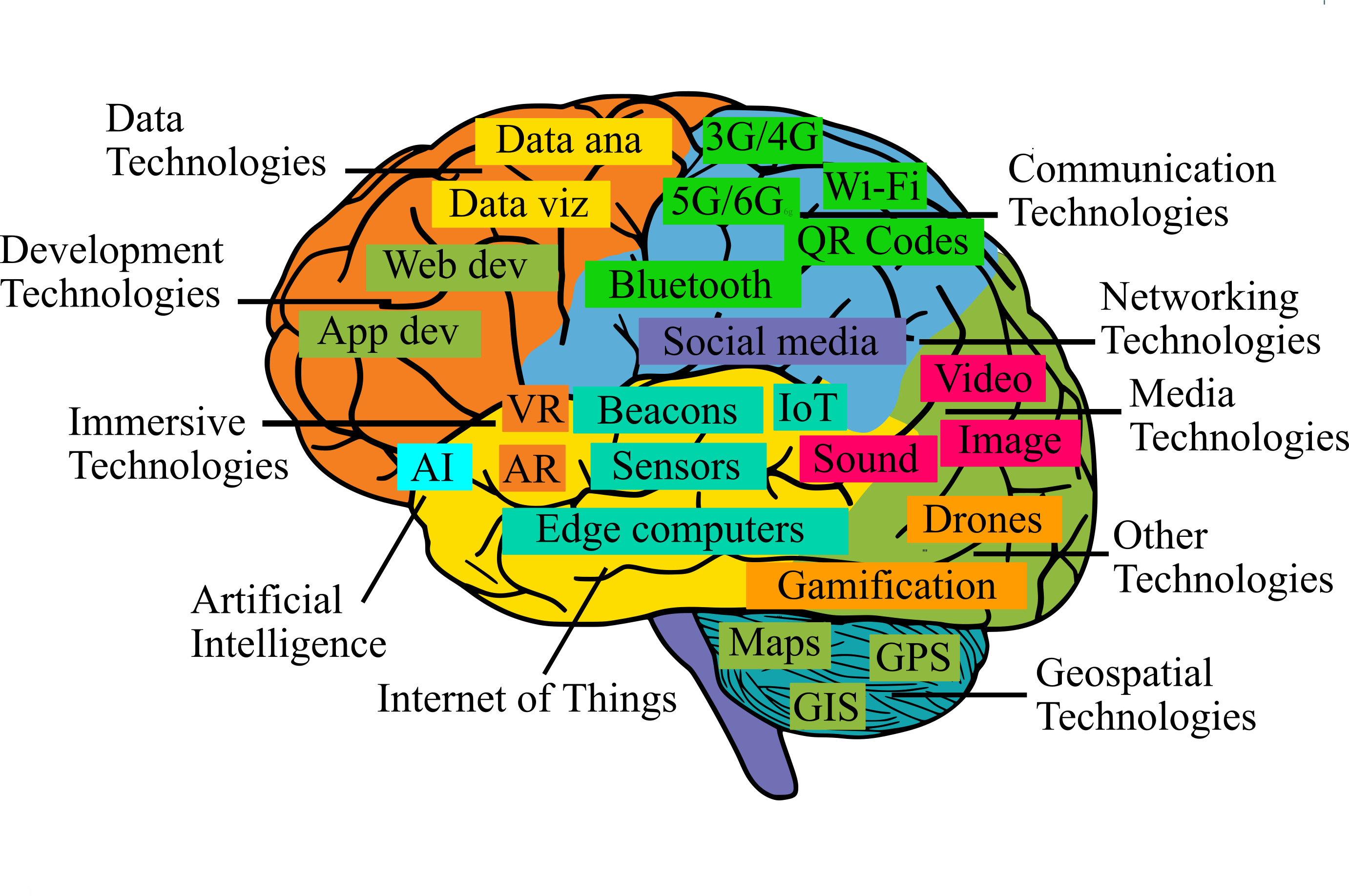}
 \caption{Technologies most commonly found in STT}
 \label{fig:STTtechs}
\end{figure}

In addition, we have compiled the technological areas that make up the tools analyzed. A non-exhaustive list is presented in Figure \ref{fig:STTtechs} that tries to map the technological areas to the cognitive areas of the human brain.

The compiled domains of action cover all application areas found both in the literature review, and in the responses to the web app form (see next section), and make up the backbone of the STT taxonomy that will integrate the European Observatory's Smartness Index.
The latter should take into account the specific goals of STT, the technological and implementation aspects, the potential impact on local (and global) sustainability, and the existence of stakeholder involvement.

\subsection{ST(T) definition}
\label{subsec:Methodology_STT_definition}
To achieve a consensual definition of STTs and create a robust taxonomy for ST that can cope with the technological evolution needed to populate the European STT Observatory, and create an index that can compare the smartness level of STT offer, we used a scientific collaborative approach, inviting experts to provide their views.

\subsubsection{Methodology- Questionnaire}

We created a form-like web app considering the following principles:

\begin{itemize}
    \item {be intuitive and easy to answer}
    \item {provide help content}
    \item {have no mandatory answers}
    \item {do not collect personal data}
    \item {provide both open and closed type questions}
    \item {be able to capture conceptual and technological aspects from the literature review}
    \item {be engaging with a game-like feel to promote answering in all fields}
    \item {provide motivational quotes as responding progresses}
    \item {reward users answering all questions with a prize: a virtual tourism experience}
    \item {allow users to leave comments and/or personal information if desired}
    \item {allow users from different countries to view and respond in their native language}
\end{itemize}

Next, we briefly present the web app GUI (Graphical User Interface) explaining how answering is achieved.

\begin{figure}[htb]
 \centering
 \includegraphics[width=\textwidth]{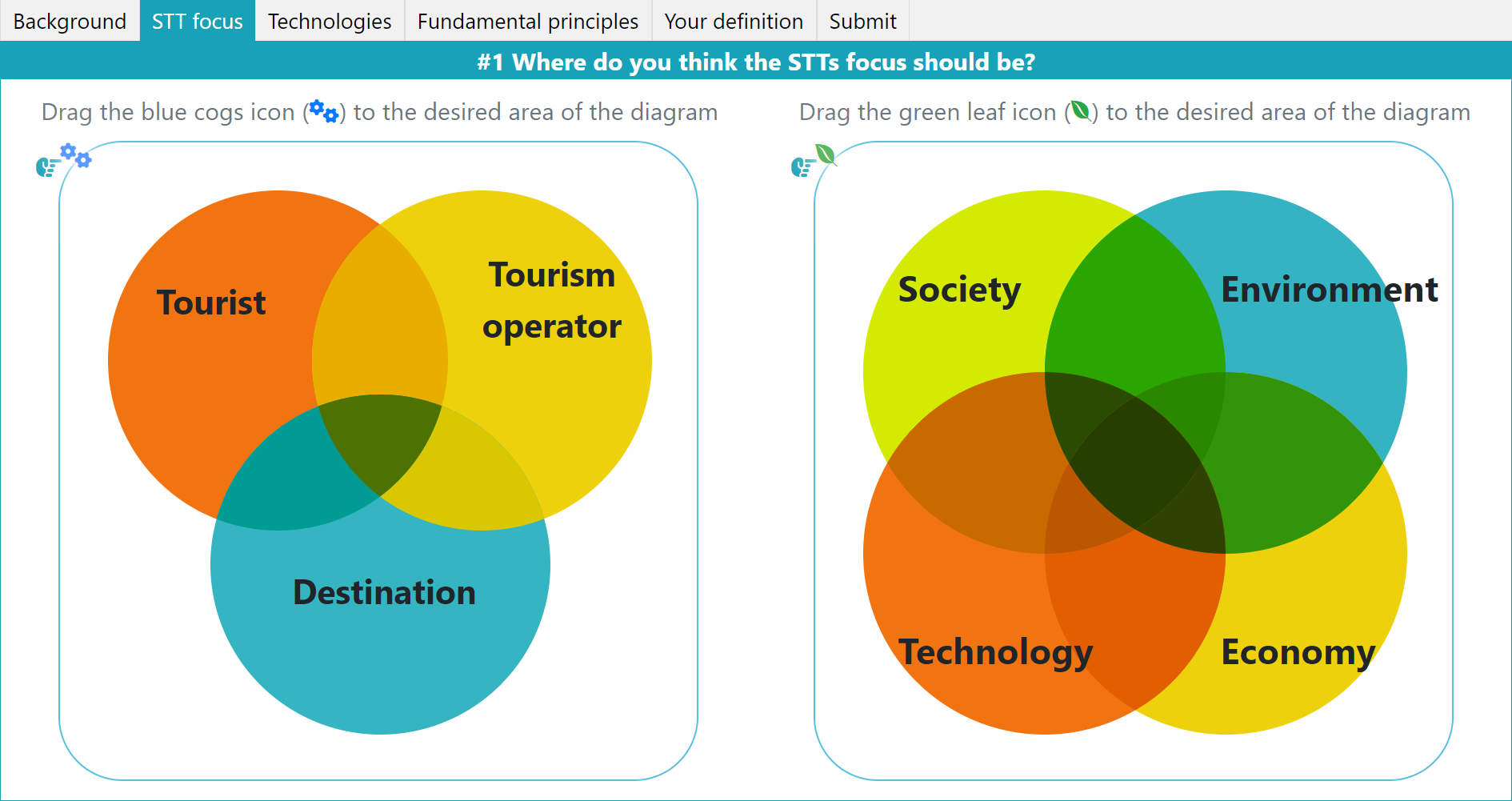}
 \caption{STT focus}
 \label{fig:STTfocus}
\end{figure}

The first question was about the focus of STT in terms of the nexus: Tourist - Destination - Tourism Operator; and Sustainability - Technology. The respondents were asked to drag the icon to the desired area in the Venn diagram (Figure \ref{fig:STTfocus}).

Regarding the nexus - Sustainability - Technology, we later identified an error in the formulation of the Venn diagram: the chosen approach does not allow users to select the individual intersections between {Society, Technology} and {Environment, Economy} that are hidden behind higher-level central intersections.

\begin{figure}[htb]
 \centering
 \includegraphics[width=\textwidth]{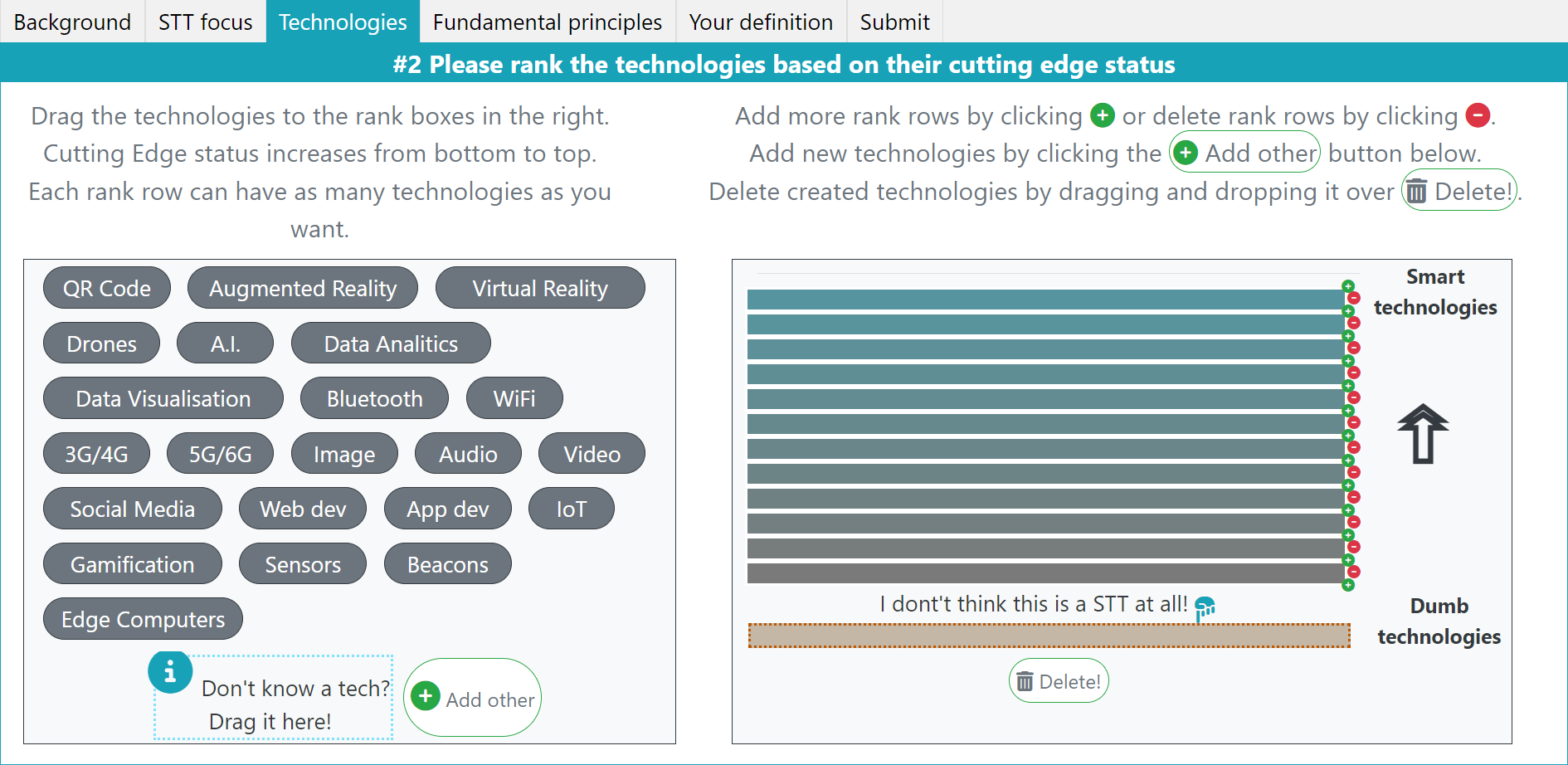}
 \caption{STT technologies}
 \label{fig:STTtechnologies}
\end{figure}

The second question on the technologies that make up STT asks users to rate the "smartness" of the technologies most commonly identified in STT offer (Figure \ref{fig:STTtechnologies}).

\begin{figure}[htb]
 \centering
 \includegraphics[width=\textwidth]{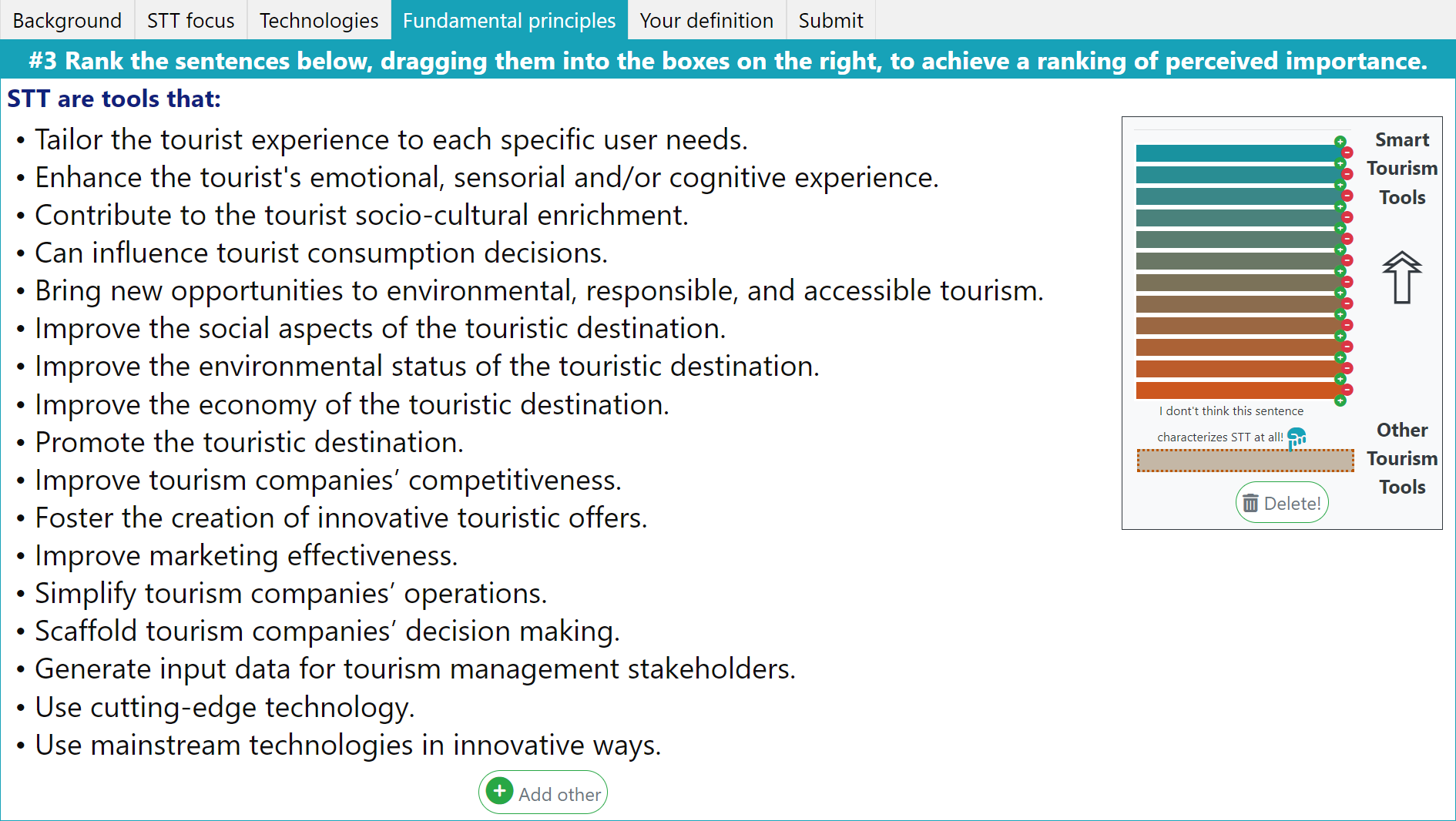}
 \caption{ST principles}
 \label{fig:STTprinciples}
\end{figure}

The third question, on the basic principles of ST (Figure \ref{fig:STTprinciples}), requires comparing different concepts underlying the definition of ST derived from the literature review.

In both questions 2 and 3, users are asked to rank the given hypotheses by dragging them into the desired box. Users can place multiple items in each rank box and add or remove rank levels at will to allow for a more discriminating separation of alternatives. In addition, users can also create and rank their own technologies and/or principles. The fourth and final question is an open-ended question where users can enter their own definition of ST(T).

\subsubsection{Methodology- Questionnaire respondents}

The population of interest for our survey is the community of ST experts. In order to maximize the chances of obtaining a large number of respondents, we searched the following databases for papers containing the term 'smart tourism' in the title, abstract and/or keywords, and for the contact details of their authors: \href{https://www.scopus.com/}{Scopus}, \href{https://link.springer.com/}{SpringerLink}, \href{https://www.webofscience.com/}{WebOfScience}, \href{https://www.lens.org/}{lens.org}, \href{https://app.dimensions.ai}{Dimensions} and \href{https://search.scielo.org/}{Scielo}.

In addition, since experts in a given field usually meet with their peers to validate their findings and to inspire themselves through discussion, we also retrieved the contacts of the authors presenting at these annual events focused on ST topics:

\begin{itemize}
  \item \href{https://enter-conference.org/}{The e-Tourism Conf. of the Int. Federation for IT and Travel \& Tourism (ENTER)}
  \item \href{https://iacudit.org/}{Int. Conf. of the Int. Association of Cultural and Digital Tourism (IACuDiT)}
  \item \href{https://www.icotts.org/}{Int. Conf. on Tourism Technology \& Systems (ICOTTS)}
\end{itemize}

We sent a personalized email to each of the identified ST experts, inviting them to participate in the survey. 

\subsubsection{Methodology- Data collection and analysis}
In the development of our web app, we opted to collect both categorical and open-ended data to allow for statistical analysis of the results aimed at identifying the most consensual technologies and principles underlying Smart Tourism Tools (STT), as well as identifying any new aspects and technologies not previously identified in the literature review. These two types of data underwent different analysis processes, which are explained in further detail in the next section.

The analysis results represent the weights of the sub-indices that will be aggregated to form the STT Smartness Index. In the following section, we present the finalized weights of the sub-indices, along with their rationale and mathematical formulation, where applicable. As the analysis is still ongoing, the final weights of all sub-indices are not yet available, as explained in the next section where appropriate.

\section{Results - presentation and analysis}
\label{subsec:PreliminaryResults}

We received 334 expert responses worldwide, with an average completion rate of 82\% (measured within the web application itself). The web application's built-in completeness checker measures the partial completeness of each question, meaning that if a respondent fails to rank a given variable in a ranking question, they will not receive a full score. Looking at completeness based on whether the question was answered or not, the average completeness rises to 91\% (Table \ref{table:formCompleteness}). Note that the average shown is based on the variables originally provided (excluding respondents' own variables) and also excludes responses to additional comments and contact information. These exclusions were also applied in the web app's completeness checker, as these variables represent additional information to the survey.

\begin{table}[htb]
\centering
\caption{Average form completeness}
\begin{tabular}{|l|c|}
\hline
\rowcolor[HTML]{ececec}
 Question & \% of responses \\ \hline
 Rank tool custom user STT definitions & 2\% \\
 \textbf{STT definition rank} & \textbf{95\%} \\
 Rank tool custom user technologies & 10\% \\ 
 \textbf{Technologies rank} & \textbf{95\%} \\
 \textbf{STT focus: tourist, destination, company} & \textbf{92\%} \\
 \textbf{STT focus: sustainability, technology} & \textbf{93\%} \\
 \textbf{Respondent STT definition} & \textbf{81\%} \\
 Respondent comments & 25\% \\
 Email contact information & 61\% \\ \hline
 \textbf{Average completeness} & \textbf{91\%} \\ \hline
\end{tabular}
\label{table:formCompleteness}
\end{table}

\begin{figure}[htb]
  \centering
  \includegraphics[width=\textwidth]{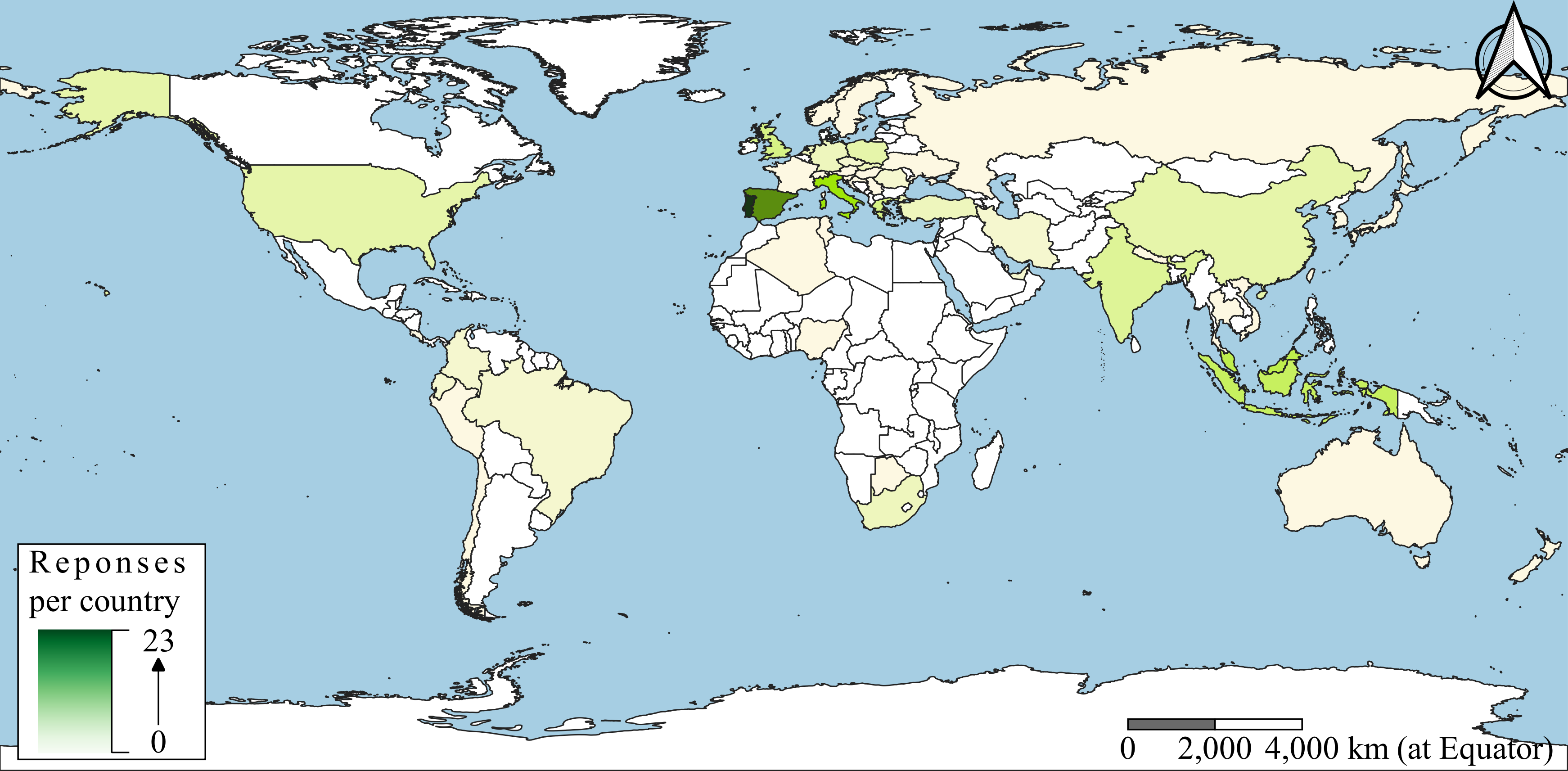}
  \caption{Worldwide distribution of responses}
  \label{fig:worldMap}
 \end{figure}

In terms of global representativeness, we received responses from 52 countries (Figure \ref{fig:worldMap}), although the true global representativeness may be higher as the geo-referencing is based on the (optional) email domains provided, meaning that only 47\% of the responses could be geo-referenced. In addition, 29 of the responses were in 10 different languages other than English - Greek, Korean, Spanish, Italian, Slovak, Chinese (simplified), Ukrainian, Portuguese, Polish and Persian.
  
\begin{figure}[htb]
    \centering
    \begin{minipage}{0.5\textwidth}
        \centering
        \includegraphics[width=0.9\textwidth]{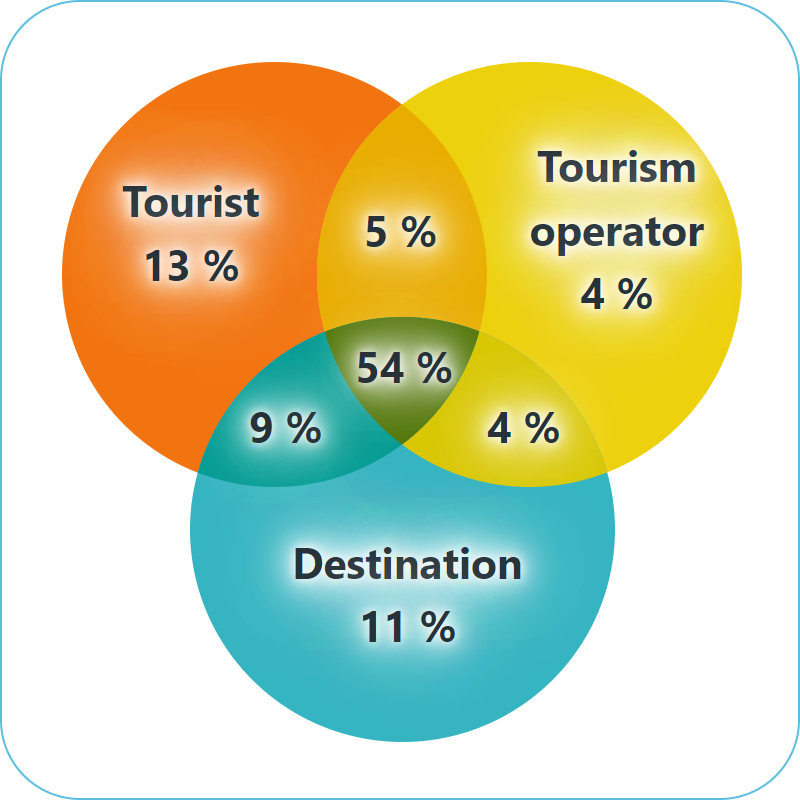}
        \caption{Tourist-Operator-Destination responses}
        \label{fig:Nexus_T-O-D_responses}
    \end{minipage}\hfill
    \begin{minipage}{0.5\textwidth}
        \centering
        \includegraphics[width=0.9\textwidth]{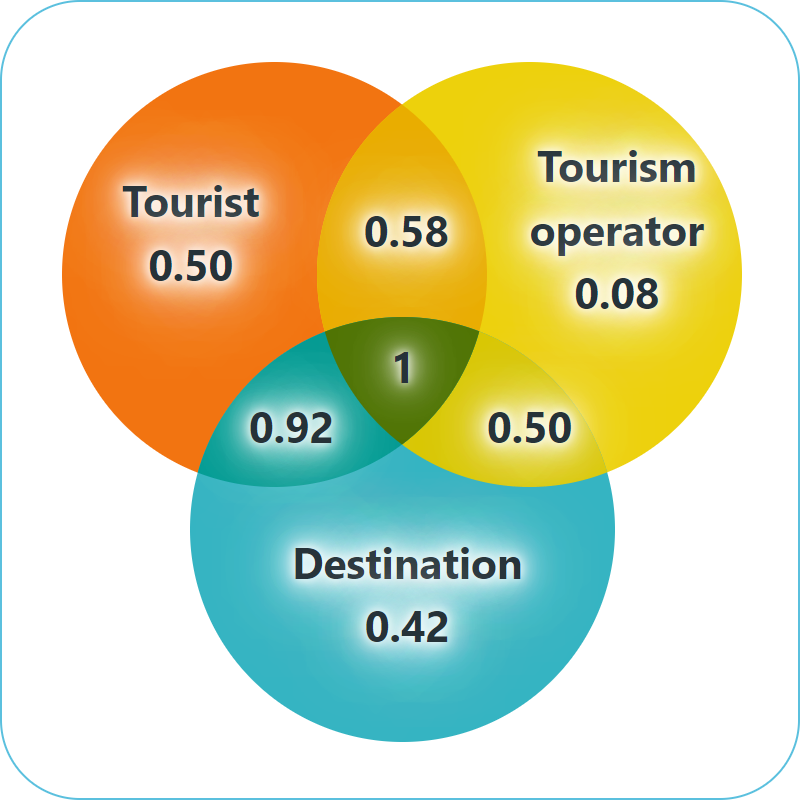}
        \caption{Tourist-Operator-Destination   results}
        \label{fig:Nexus_T-O-D_results}
    \end{minipage}
\end{figure}

Figure \ref{fig:Nexus_T-O-D_responses} shows the results of the answers to first question, concerning the focus of STT regarding the nexus - Tourist-Operator-Destination.


Visual analysis of the diagram reveals a consensus of the STT focus on the Tourist and the Destination to the detriment of the Tourism operator. Further analysis of the diagram seems to indicate that the users' perception of the data collection method was not consistent across the universe of STT experts.  As the data collection approach is innovative, this may be the case. It is reasonable to assume that the problem under study should be additive, i.e. the intersections of the groups should have higher values (or at least the same order of magnitude) than the independent groups, i.e. if one group, let's say "Tourist", is key to STT, then intersections containing "Tourist" should retain at least some of that importance and add the importance of the intersecting group, even though the importance of each independent group in the intersected group might be different from their importance in the independent groups. To overcome this and perform a deeper analysis of the data, and ultimately extract indicators for the Smart Index, we analyzed the data considering each level of intersection as a separate universe. In this way, we can derive the relative importance of each of the nexus elements within each intersection universe.

Let’s assume that participants in the dual intersection universe assign an intrinsic importance or weight to each of the base variables - Tourist, Destination, Company - when responding. This would mean that we can calculate an average weight or ponderation for each of these variables within the dual intersection universe.
This can be expressed as:

\begin{equation}
Result_i = \alpha Tourist + \beta Destination + \gamma Company 
\label{eq:Venn1BaseEq}
\end{equation}
where:
\begin{description}
  \item[i] is the group of all possible intersections in that universe. In the case of the dual intersection universe - [Destination\_Tourist, Tourist\_Company, Company\_Destination],
  
  \item[$\alpha$, $\beta$, and $\gamma$] represent the weights of the base variables - Tourist, Destination, Company -, constant within the intersection universe.
\end{description}

Using equation \ref{eq:Venn1BaseEq} we can create a system of equations for each of the groups in a given universe. Note that for the dual intersection universe eq.1 only has two terms.
We can then create a matrix for the equation system - A - taking '½' for each of the base variables present in each dual intersection world, as we are only analyzing dual intersections.

\begin{equation}
A \cdot X = B\\
\begin{bmatrix}
0.5 & 0.5 & 0 \\
0.5 & 0 & 0.5 \\
0 & 0.5 & 0.5
\end{bmatrix}
\cdot
\begin{bmatrix}
\alpha \\
\beta \\
\gamma
\end{bmatrix}
=
\begin{bmatrix}
0.50 \\
0.29 \\
0.21
\end{bmatrix}
\end{equation}

where:

\begin{description}
  \item[A] is the systems equation matrix,
  
  \item[X] is the matrix of weights $\alpha$, $\beta$, and $\gamma$, that we are trying to find, and

  \item[B] is the matrix containing the ratio of dual intersection i, over the total of the dual intersections universe, calculated directly from the gathered data.
\end{description}

We can then solve for X:

\begin{equation}
X = A^{-1} \cdot B
\end{equation}

where:

\begin{description}
  \item[$A^{-1}$] is the inverted matrix for the equation system.
\end{description}

X contains the average intrinsic weights of the universe of the double intersection. We can directly estimate the average intrinsic weights for the independent universe by calculating the ratio of the entries in each independent group to the total entries for the independent universe. The average weights of the whole universe - independent \& intersecting - can then be estimated using a weighted average, taking into account the dimensions of each universe in the data collected.

\begin{table}[htb]
\centering
\caption{Average weights Tourist-Operator-Destination nexus}
\begin{tabular}{|p{0.12\linewidth} | p{0.12\linewidth} | p{0.12\linewidth}|}
\hline
\rowcolor[HTML]{ececec}
 Tourist & Destination & Operator \\ \hline
 0.50 & 0.42 & 0.08 \\ \hline
\end{tabular}
\label{table:avWeightsT-O-D}
\end{table}

Table \ref{table:avWeightsT-O-D} shows the resulting overall weights for the Tourist-Operator-Destination nexus. It is clear, as identified in the visual analyses, that the Tourist is the most important focus for STT, closely followed by the Destination, and the Tour Operator (Company) is not as central to STT. We can then apply these averages to each group by adding the weights of the variables present in each group.

Figure \ref{fig:Nexus_T-O-D_results} presents the results in a Venn diagram to facilitate comparison with the data collected. As you can see, the relationships of the original data remain, but now we have the cumulative effect required by raising the levels of intersection.

Regarding the question concerning the Sustainability-Technology nexus, the approach used to collect data in this particular Venn diagram does not allow the user to select the intersections between {Society, Technology} and {Environment, Economy}. However, it is important to note that although the user cannot enter data at the two intersections mentioned, the graphical nature of the response method leads the user to spatially locate their responses in the regions closest to their intended focus. This means that it is still possible to draw valid conclusions from the collected data.

\begin{figure}[htb]
    \centering
    \begin{minipage}{0.5\textwidth}
        \centering
        \includegraphics[width=0.9\textwidth]{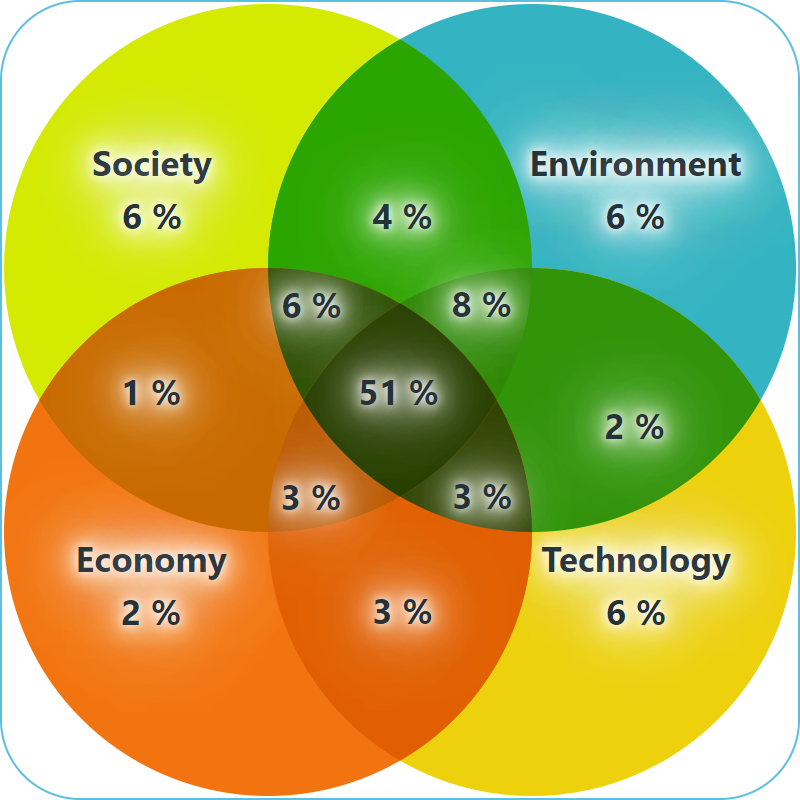}
        \caption{Sustainability-Technology responses}
        \label{fig:Nexus_S-T_responses}
    \end{minipage}\hfill
    \begin{minipage}{0.5\textwidth}
        \centering
        \includegraphics[width=0.9\textwidth]{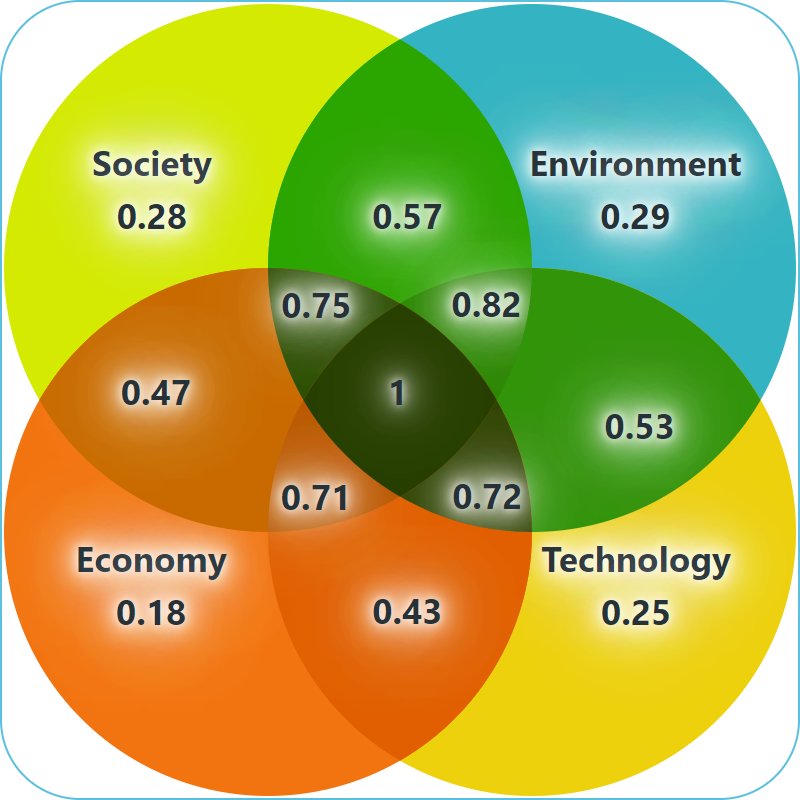}
        \caption{Sustainability-Technology results}
        \label{fig:Nexus_S-T_results}
    \end{minipage}
\end{figure}

A visual analysis of this diagram (Figure \ref{fig:Nexus_S-T_responses}) shows that the focus is on the environment, society, and technology, and their combinations, to the detriment of the economy. Two interesting conclusions can be drawn from this. Firstly, although the term "smart" is intertwined with technology, at least in the field of SST, the emphasis seems to be primarily on the environment and society, and only then on technology. Secondly, technology seems to bypass one of the pillars of the tripartite concept of sustainability, the economy.

In this case, because we have the data gaps mentioned above, we chose a simpler approach. We decided to treat the data as a whole, so we created an overall average by multiple counting the overlapping data, thus creating a new universe where our sample data is 3 times larger (in the largest dataset of tested alternatives) compared to the collected data. We tested 6 hypotheses for extracting the average weights with different combinations of:

\begin{itemize}
    \item Considering or not the central intersection - to test the effect of the largest group (about 50\% of the total responses, when all other groups exist)
    \item Considering or not double intersections, replacing missing values with:
    
    \begin{itemize}
        \item the average size of the remaining universe of double intersections,
        \item an average for the given variables within the remaining universe of double intersections
    \end{itemize}
\end{itemize}

The alternative that best fits the data does not consider the central intersection and considers the double intersections, replacing the missing values with an average for the variables in question in the remaining universe of double intersections. The results are shown in Figure \ref{fig:Nexus_S-T_results}.

Next, we analyze the results of the second question, where respondents were asked to rank a list of technologies according to their state of the art. In addition, users could suggest and rank their own technology proposals. We received 72 technology suggestions, of which 49 were unique and some were subsets of our own list.

\begin{figure}
  \centering
  \begin{subfigure}{0.49\textwidth}
      \includegraphics[width=\textwidth]{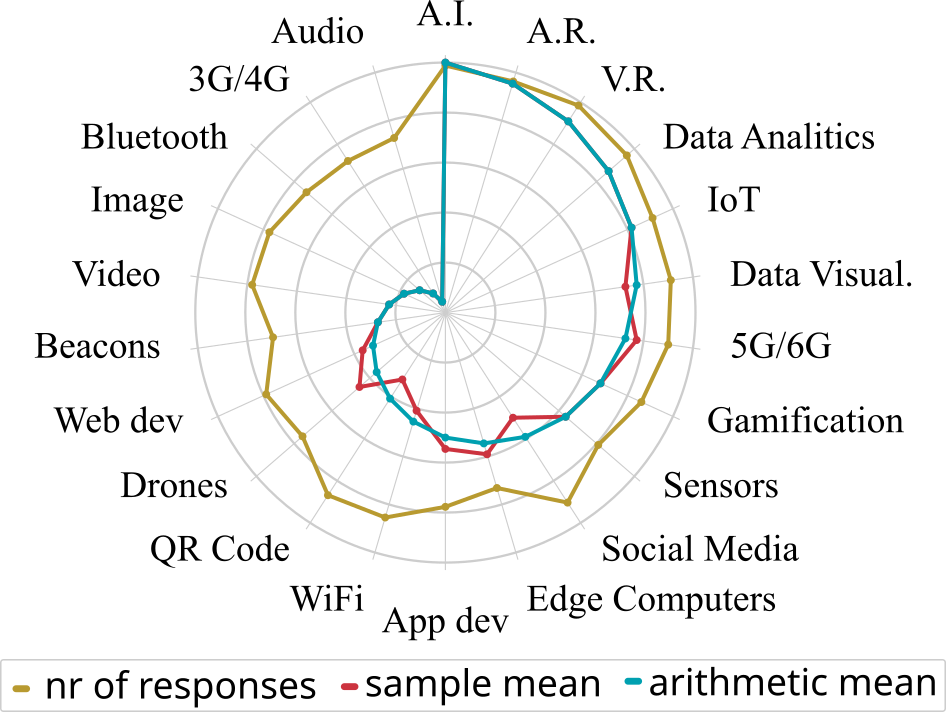}
      \caption{linear ranking order}
      \label{fig:tech_rank_a}
  \end{subfigure}
  \hfill
  \begin{subfigure}{0.49\textwidth}
      \includegraphics[width=\textwidth]{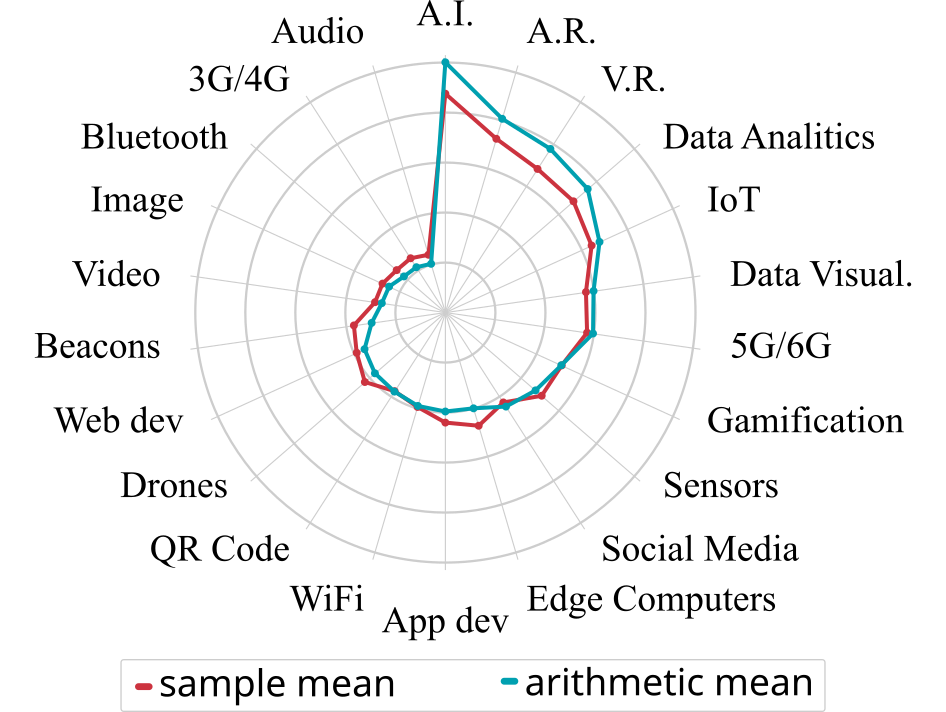}
      \caption{calculated means}
      \label{fig:tech_rank_b}
  \end{subfigure}
  \caption{ Average ranks found.}
  \label{fig:tech_ranks}
\end{figure}

 Concerning the ranking of the technologies, excluding the custom ones in this analysis, we received answers where users did not classify some technologies, although there is a box to drop technologies not considered as STT. We cannot be sure if technologies not ranked were considered not STT or if this was just an issue of incomplete answer, or even a combination of the former two situations. To test the influence on the results, we created two separate average ranks, one considering the arithmetic mean and another considering the sample mean (Figure~\ref{fig:tech_ranks}). Figure \ref{fig:tech_rank_a}) shows the ranking in descending order of cutting-edge status, from left to right, taking into account the arithmetic mean rank. It does not show arithmetic averages, but the ranking at equal intervals. In this way, a perfect blue spiral line is obtained, reflecting the steady descent of the ranking.

The number of ranked technologies is different for each technology (brown line in Figure \ref{fig:tech_rank_a}), reflecting the different sample sizes per technology that influence the results. The behavior of the red line follows a spiral in the higher and lower rankings, but in the intermediate rankings, there are several disagreements between statistical approaches. If we analyze the average rank expressed in terms of the arithmetic mean actually calculated (Figure \ref{fig:tech_rank_b}), the differences diminish, although they are still present. This figure expresses the actual distances between the technologies proposed by the respondents and shows that the perceived technological status is close for the divergent ranking categories. This approach also highlights the positive distances between adjacent technologies, e.g. the leap from AI to AR is much greater than the leap from the latter to VR, and the same effect can be seen between other technologies, although not as pronounced. 

In response to the question which asked users to rank some of the fundamental principles found in the literature review and to provide their own fundamental principles, we received 13 principle suggestions from 8 users (Table \ref{table:customPrin}). These propositions reveal important aspects such as tourist mobility or safety.

\begin{table}[htb]
\centering
\caption{Fundamental principles proposed by respondents}
  \begin{tabular}{|l|l|}
  \hline
  \rowcolor[HTML]{ececec} 
  \textbf{User\_ID}                             & \textbf{Proposed fundamental   principles}                                   \\ \hline
                          
  32                                            & Help tourism destinaiton   to be more sustainable, competitive and resilient \\ \hline
                          
                                               & Improves   mobility efficiency                                               \\ \cline{2-2} 
                          
  \multirow{-2}{*}{47}  & Enhance the tourist's   experience                                           \\ \hline
                          
  71                                            & Metaverse                                                                    \\ \hline
                          
  96                                            & Creating   Tourism Network                                                   \\ \hline
                          
  118                                           & Optimize   the tourist route                                                 \\ \hline
                          
  128                                           & Bring   better experience through smart chain management                     \\ \hline
                          
                                               & It   supports the improvement of tourism destination branding                \\ \cline{2-2} 
                          
  \multirow{-2}{*}{250} & It supports the   improvement of users' Customer Satisfaction                \\ \hline
                          
                                               & Promote   collaboration among the stakeholders of a tourist destination      \\ \cline{2-2} 
                          
                                               & Strengthen the governance   of the destination                               \\ \cline{2-2} 
                          
                                               & Facilitating tourist   mobility around the destination                       \\ \cline{2-2} 
                          
  \multirow{-4}{*}{318} & Increasing security in a   tourist destination                               \\ \hline
  \end{tabular}
  \label{table:customPrin}
  \end{table}

\begin{figure}[htb]
 \centering
 \includegraphics[width=\textwidth]{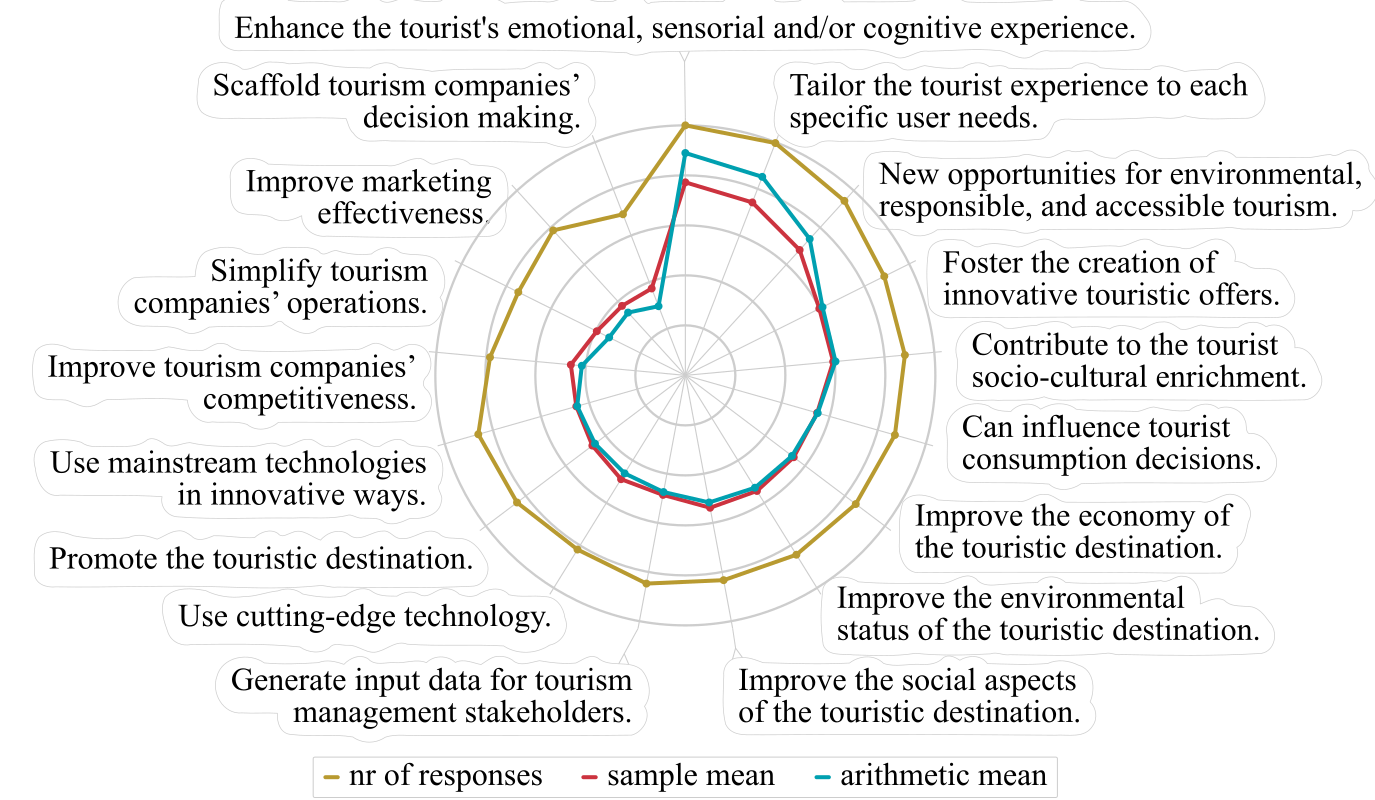}
 \caption{STT fundamental principles average ranks}
 \label{fig:principleRank}
\end{figure}

  Regarding the ranking of fundamental principles provided in the web app, we followed the same strategy as described above for the technology ranking, since some principles were not ranked. In this case, although three fundamental principles change places between the two statistical approaches (although they have similar ranks), the experts were very consensual in their ranking of the fundamental principles of STT (Figure  \ref{fig:principleRank}).

Interestingly, the results are very similar to those obtained in the first question, on the focus of STT. The tourist and the tourist experience come first, followed by the destination and environmental concerns, and then technological considerations, followed by commercial aspects.

For both questions regarding the ranking of data, we have chosen to use the sample average to construct the Smart Index because we believe it is the most appropriate statistical approach to construct the ranking of ST fundamental principles, as it takes into account the variability and richness of the data collected. This is due to the fact that:

\begin{itemize}
    \item The sample size is relatively large (334 responses from experts around the world) and representative of different regions and perspectives
    \item The design of the survey was intuitive, easy, engaging and rewarding, which could have reduced the likelihood of non-response and incomplete answers
    \item The survey used a rating tool that allowed multiple items in each rating box and the addition or deletion of rating levels at will, which could reflect the experts' preferences and opinions more accurately
\end{itemize}

\begin{figure}[htb]
    \centering
    \begin{minipage}{0.49\textwidth}
        \centering
        \includegraphics[scale=0.08]{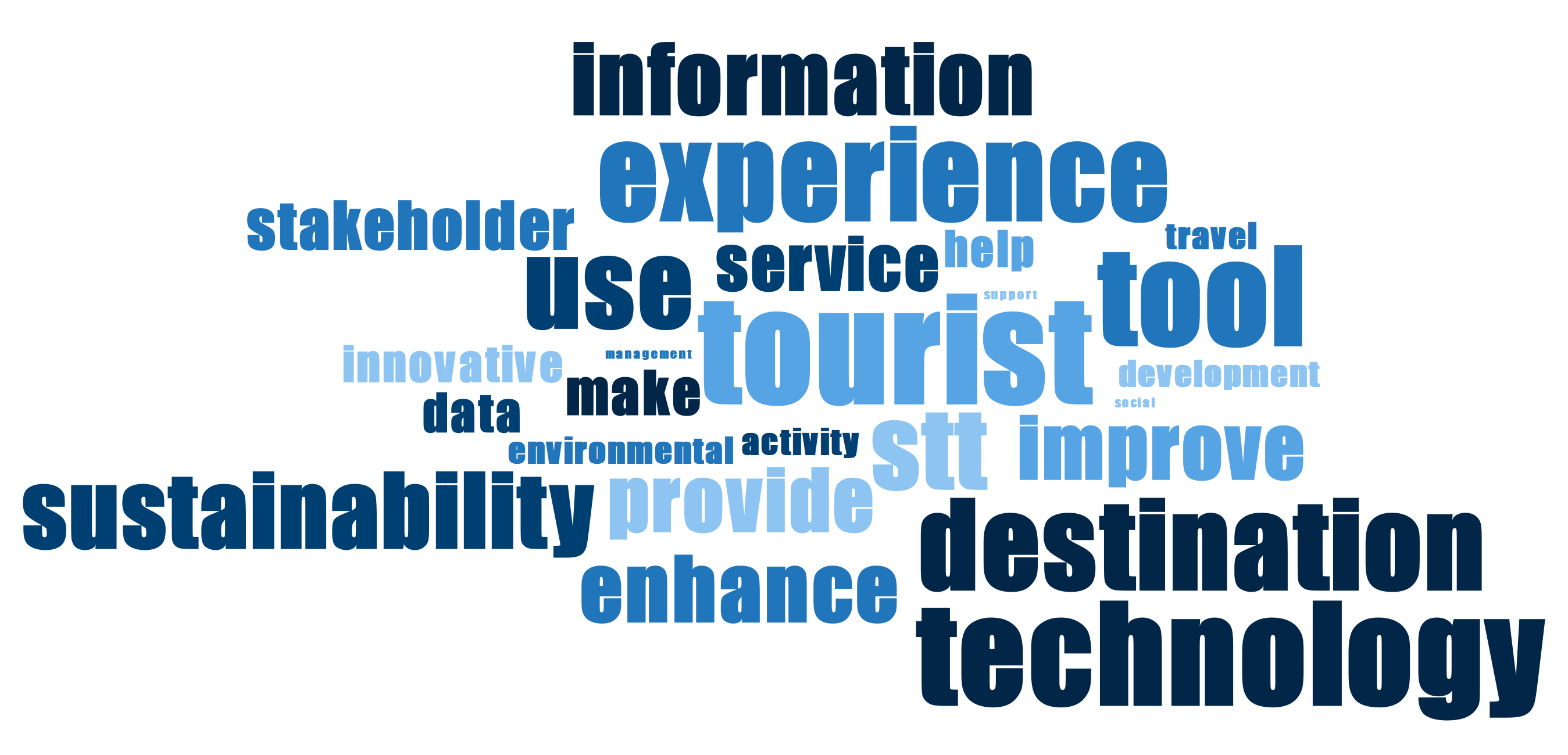}
        \caption{Single word cloud}
        \label{fig:OneWordClowd}
    \end{minipage}\hfill
    \begin{minipage}{0.49\textwidth}
        \centering
        \includegraphics[scale=0.08]{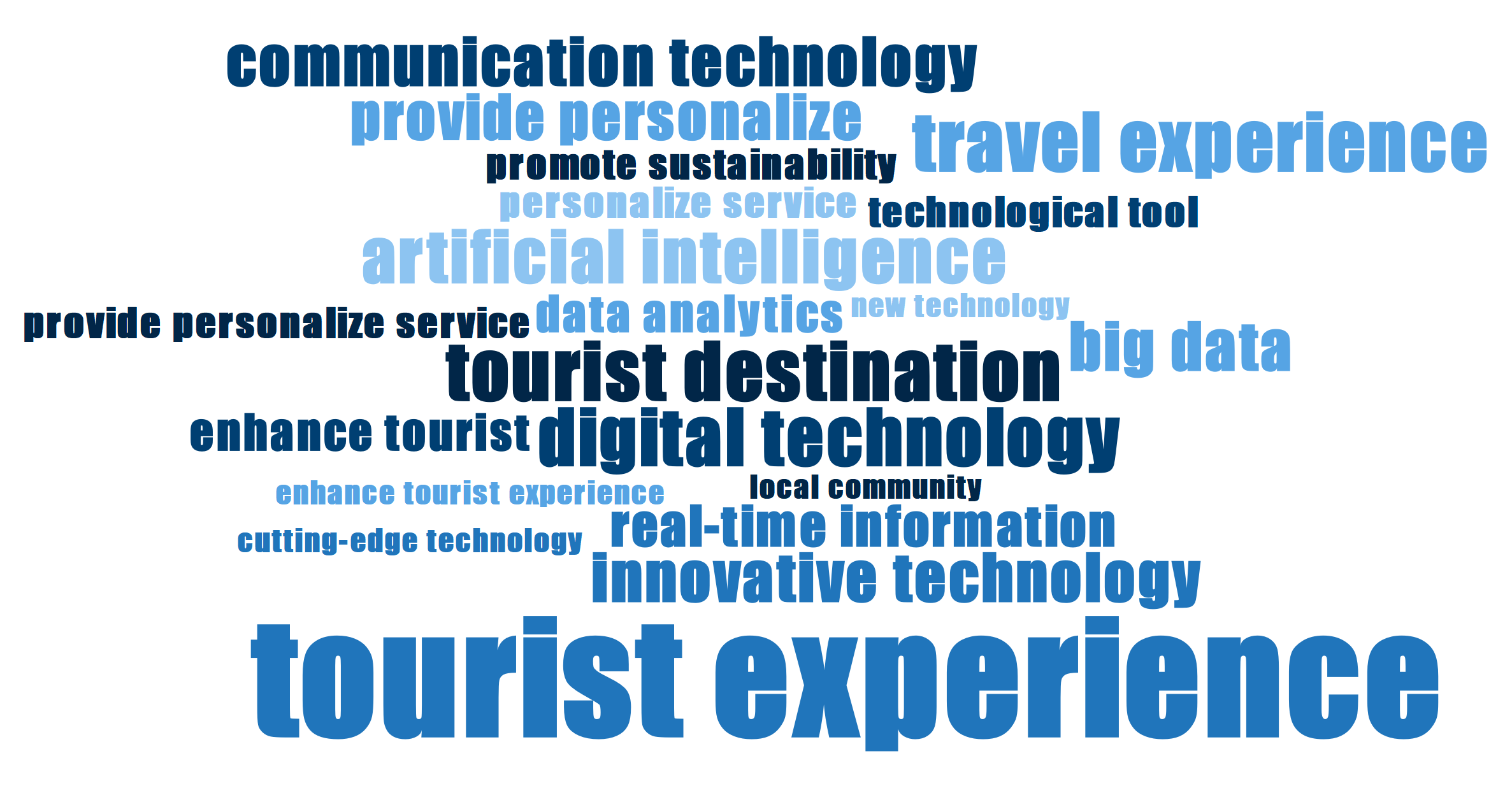}
        \caption{Bigram and trigram word cloud}
        \label{fig:TwoWordClowd}
    \end{minipage}
\end{figure}

Finally, users could write their own definition of STT in an open-ended question. We first used MAXQDA5\footnote{\url{https://www.maxqda.com/}} to help analyze the responses. We started by creating two-word clouds, one with just single words (Figure \ref{fig:OneWordClowd}) and the other with bigrams and trigrams (Figure \ref{fig:TwoWordClowd}).

Once again we can see the importance of the tourist, destination, and sustainability in the definitions presented. We then coded the responses according to three main categories, (i) STT objectives, (ii) STT target categories, and (iii) technology areas.

\begin{figure}[htb]
  \centering
\begin{minipage}{0.5\textwidth}
      \centering
      \includegraphics[width=\textwidth]{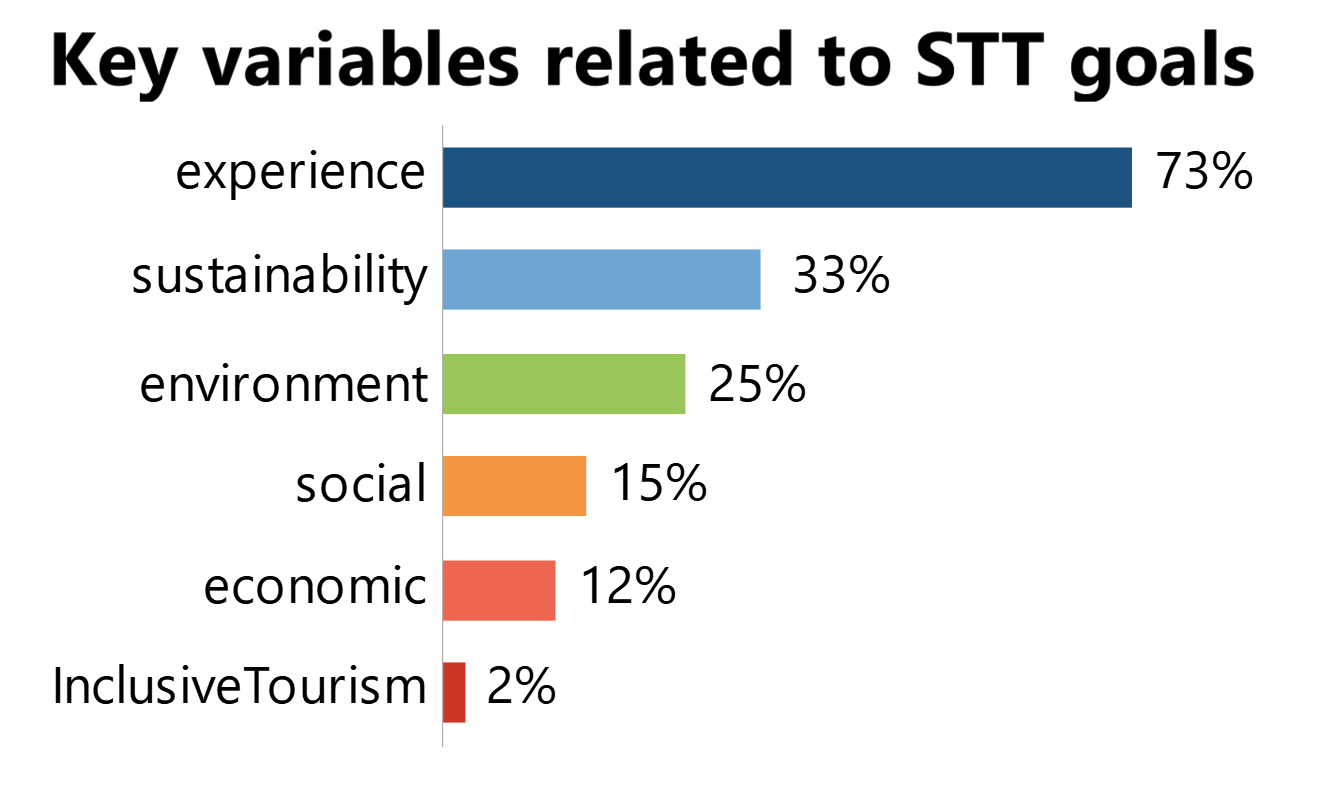}
      \caption{Code frequencies: STT goals}
      \label{fig:open_vars_defin}
  \end{minipage}\hfill
\begin{minipage}{0.5\textwidth}
      \centering
      \includegraphics[width=\textwidth]{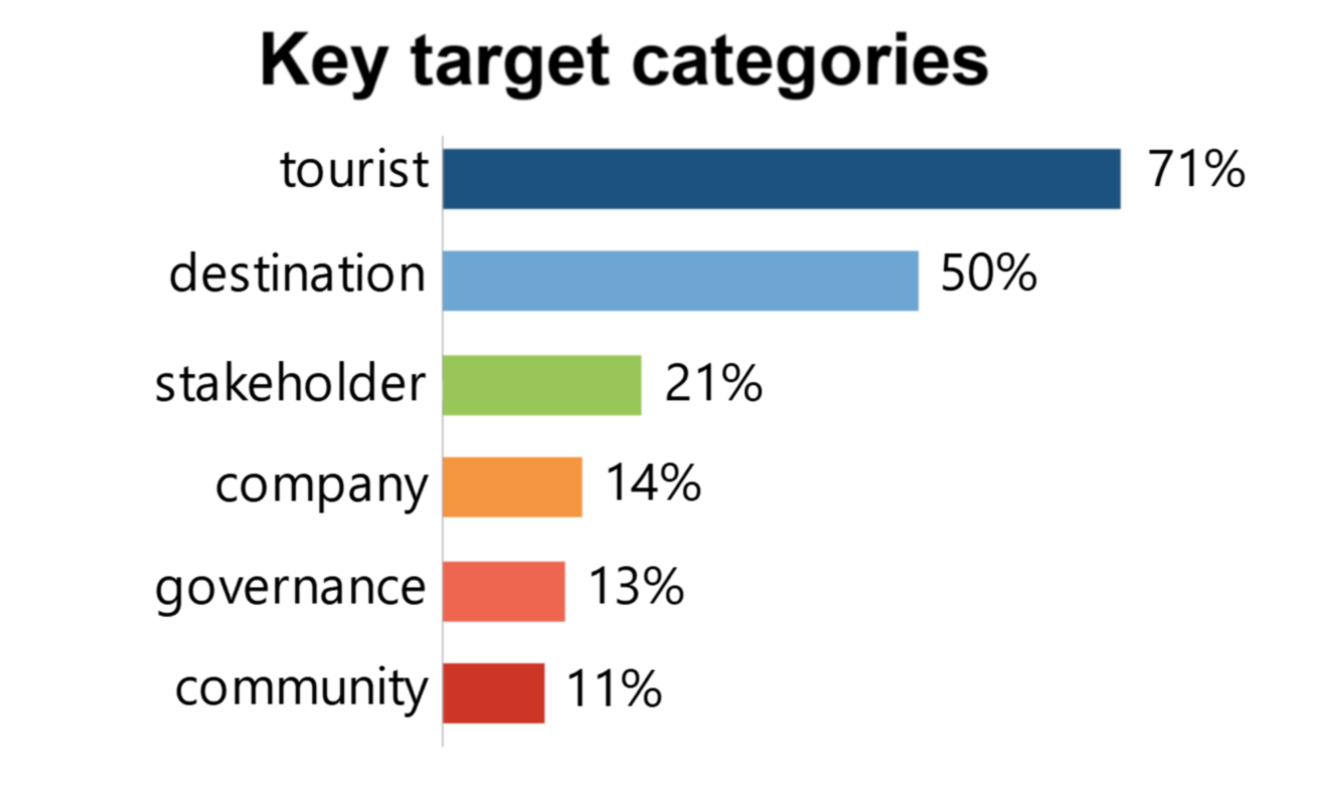}
      \caption{Code frequencies: STT target categories}
      \label{fig:open_cats_defin}
  \end{minipage}
\end{figure}

The code frequency analysis in relation to the STT objectives is shown in Figure \ref{fig:open_vars_defin}. Again we see the same pattern as in the previous responses. It is worth noting that some mentions of environmental aspects could be related to broader sustainability, as these two concepts are often used interchangeably, and this was evident in our context analysis. The same could be said for inclusive tourism. The code for this category only highlights responses that specifically mention it, but other mentions of the social aspects of sustainability could also consider this aspect.

The analysis of the code frequency regarding the key target categories is presented in Figure \ref{fig:open_cats_defin} and again shows the same patterns as previously observed in relation to the STT definition. It should be emphasized that the \textbf{stakeholder} code represents the mentions where respondents broadly referred to all direct and indirect stakeholders in the tourism industry.

\begin{figure}[htb]
 \centering
 \includegraphics[scale=0.11]{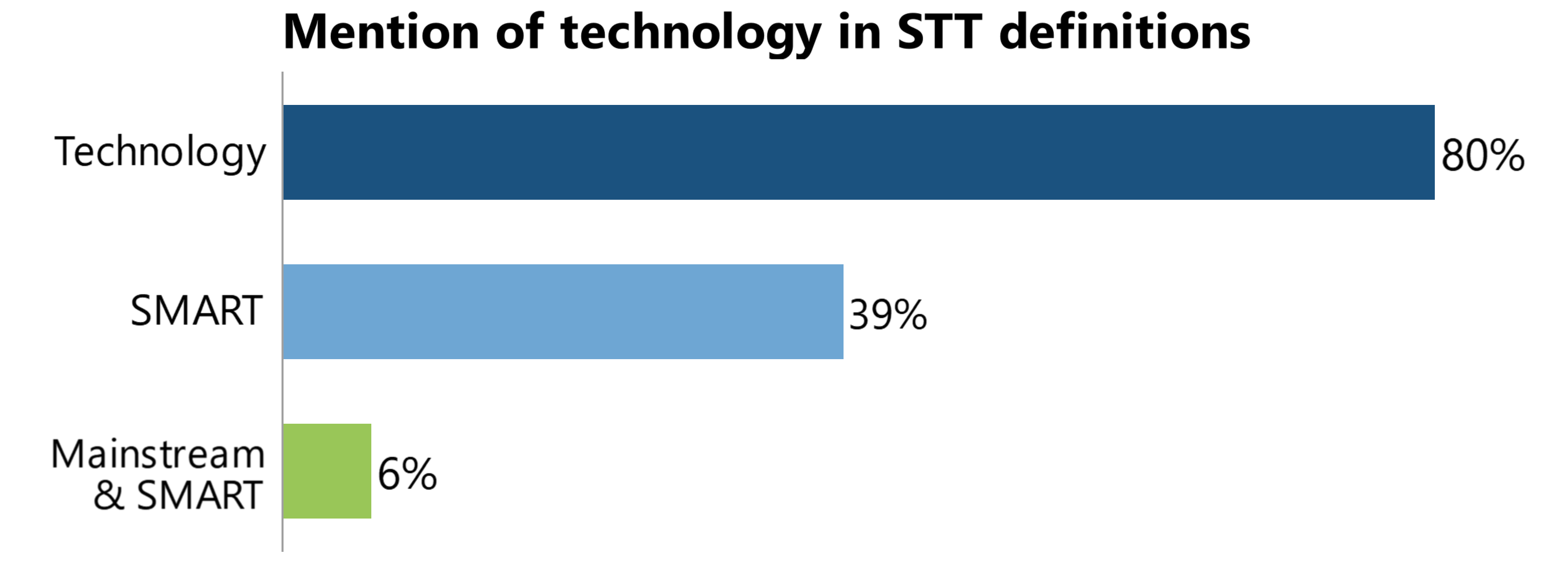}
 \caption{Code frequencies regarding technology mentions}
 \label{fig:technologyMentions}
\end{figure}

The code frequencies for technology mentions are shown in Figure \ref{fig:technologyMentions}. The code \textbf{Technology} refers to mentions of technology as a broad field and does not distinguish between smart and mainstream technologies, i.e. these could refer to combinations of mainstream and smart technologies, the context analysis points in this direction. The \textbf{SMART} code aggregates mentions in different formats such as 'innovative', 'advanced', 'new', or 'cutting edge'. In the code \textbf{Mainstream SMART}, respondents specifically mentioned the use of mainstream and/or advanced technologies.

The presented approach for the analysis of the open-ened answers does not allow extrapolating meaningful indices for the Observatory smartness index. To overcome this, we tested different preprocessing techniques and Natural Language Processing (NLP) analyses. These ranged from simple approaches such as assessing text complexity, to more complex methods like Topic Modelling, a technique that identifies topics in a set of documents. We also attempted to measure data similarity in vector space, a mathematical model where words are represented as vectors in multi-dimensional space, using embeddings from small language models, simple and/or small AI models trained on text data, useful for several NLP tasks, but not highly effective on capturing context and concepts. This latter approach was also combined with a semantic search of previously extracted concepts. However, these attempts fell short of achieving the desired indices. This was largely due to the high text complexity of the definitions and their varied sentence structures. The definitions also include many examples, which add redundant or extraneous information to the statistical analysis. Since LLMs (Large Languge Models, AI models trained on vast amounts of text data, able to generate human-like text) can understand both concepts and context, they are particularly useful for treating data with such characteristics, given adequate instructions and examples. We are currently using LLMs to do Text Classification, Relationship Extraction, and Span Categorization. This approach should allow deriving a single vector of the relevant concepts for each definition that can be used to statistically determine the required indices.

\section{Final remarks}
\label{sec:Conclusion}

ST refers to the use of technology and data-based solutions to improve the tourist experience, destination management, and sustainability. It involves the collaboration of different stakeholders, such as tourists, suppliers, governments, and residents, to create value and benefits for all. ST is inspired by the concept of smart cities but also takes into account the specific needs and characteristics of tourism and tourism destinations.
STTs are the set of tools that support the development and implementation of ST. Therefore STTs are a set of tools that:

\begin{enumerate}
    \item \emph{apply the fundamental principles of ST, to achieve (at least some) ST goals,}
    \item \emph{within the specific scope envisioned by developers (application domain and specificities of the final consumer or area, ensuring concordance with ST scope),}
    \item \emph{considering the inputs of/and the final influence on the key stakeholders,}
    \item \emph{taking advantage of the relationship of their specific field, the ST field (as most STT developers are not specifically working with ST, but are experts in their own field, e.g. VR), and with other interconnected knowledge areas, either technological or not, while applying techniques and technologies and,}
    \item \emph{trying to minimize biases at an operational level (e.g. avoid search engine ”commercial” algorithm influence in AI-based tourism recommendation systems) and on a conceptual level, ensuring they are indeed STT and not just digital tools.}
\end{enumerate}

Our results, based on the experts' opinions, corroborate that Smart Tourism's definition cannot consider just a dimension, such as the technocentric one. The results further allow us to define to what degree each key aspect of the definition should be considered when evaluating Smart Tourism offer. Our STT definition cannot yet deal with the fast pace of technological evolution. Further research is required considering different approaches, such as using the \href{https://www.weforum.org/reports}{reports published yearly by World Economic Forum}, to identify and extract emerging technologies, and the number of scientific publications referring to these technologies, in the field of Smart Tourism, as a proxy for the technology Smartness. Additionally, this smart-metric approach means that, as technology evolves, earlier technologies do not become dumb, but actually less Smart, and this can be quantified.

Our proposed consensual "current" definition for STT is that of:
\begin{mdframed}
  \item \emph{seamlessly interconnected digital tools designed to benefit all stakeholders in the tourism industry, with a special focus on the tourist and the destination, that aim at local, and possibly global, sustainable development.}
\end{mdframed}
\vspace{0.5cm}
At their highest level, STTs produce experiential ecstasy in the dynamic human interface with technology, enriching and not alienating people. At decreasing smartness levels they also aim to improve the efficiency, quality, and competitiveness of tourism destinations, operations, products, and companies. They also aim to improve the quality of life of local communities and the well-being of tourists, while minimizing the negative impacts of tourism on the environment and local culture.
The level of STT Smartness depends more on achieving the objectives of sustainable tourism than on their technological status. Indeed, emerging technologies offer unforeseen possibilities, but so does innovation combined with conventional technology. This means that the intelligence of these tools should give higher importance to their alignment with the principles and objectives of Smart Sustainable Tourism. A full evaluation would require assessing tool deployment results, i.e. their impact on all relevant stakeholders in the tourism industry and on local (and possibly global) sustainability. This more complete Smartness assessment is beyond the scope of our current work since the latter only considers the potential impact of STTs based on the information available, mainly in the product descriptions and then in more detailed descriptions provided by the STT developers/owners.

\subsubsection{Future developments}
We anticipate broadening the audience for deriving the ST(T) definition to other key stakeholders, such as tourists, tourism operators, and destination managers. This will allow us to derive a more representative STT Smartness Index. Finally, we plan to use the STT Smartness Index to assess the STT offer in the European market.

\begin{acknowledgement}
This work was developed in the scope of the RESETTING project, funded by the COSME Programme (EISMEA) of the European Union under grant agreement No.101038190.\\
We also thank all survey participants without which this study would not have been possible.
\end{acknowledgement}

\bibliographystyle{styles/spmpsci}

\bibliography{MAIN}

\end{document}